\documentclass[botnum, fleqn]{unmeethesis}

\usepackage{graphicx}
\usepackage{hyperref}

\begin{document}
\ifx\href\undefined\else\hypersetup{linktocpage=true}\fi

\frontmatter

\title{Model And Fabrication Of A Proof-Of-Concept Polarimeter-In-A-Pixel}

\author{Mario A.~Serna Jr}

\degreesubject{M.S., Electrical and Computer Engineering}

\degree{Master of Science \\ Electrical Engineering}

\documenttype{Thesis}

\previousdegrees{B.S., Mathematics and Physics, US Air Force Academy, 1997 \\
                 M.S., Physics, Massachusetts Institute of Technology, 1999}

\date{August 2003}

\maketitle


\begin{dedication}
   To my wife, Laura, for her love, support, tolerance and encouragement. \\[3ex]
\end{dedication}

\begin{acknowledgments}
   \vspace{1.1in}
   I would like to thank  Dave Cardimona at the Air Force Research Laboratory, Space
   Vehicles Directorate for his support and funding, and my advisor, Dr.~Sanjay Krishna,
   for his support and advice.
   I owe a big thank-you to Peter Hill, Sunil Raghavan, Beth Fuchs, and John Zou for all their help with processing
   the sample.
   I need to thank Z.~Liau at MIT Lincoln Laboratory for help with
   wafer fusion, and Kathy Sun and Sanh Luong at Zia Laser for help with lapping and polishing the device.
   I also would like to thank Paul Rotella
   Greg von Winckel, and Joe Evans Jr for their help designing, ordering, growing, and processing
   aspects of the demonstration polarimeter.
   For help with characterization of the various test devices, I need to thank
   Dang Le, Scott Gingrich, Ryan McGuire, and Beverly Klemme.
   Last, I would like to thank Dan Huang, Chris Morath, Paul Alsing, Tzveta
   Apostolova, Paul LeVan, Tom Caudill, Scott Tyo, J Piprek,  Susumu Noda, Lutz Wendler  and Shawn Lin for many hours
   of fun and helpful discussions.  I appreciate Barrett Flake's
   help with computing resources on several weekends.  I also
   appreciate Michael O'Brien and Arezou Khoshakhlagh
   for their help with SEM and AFM images.
\end{acknowledgments}

\maketitleabstract

\begin{abstract}
I propose a new optoelectronic device that completely and
instantaneously measures the incident light's polarization for a
narrow wavelength band in a single physical pixel. The device has
four (or more) quantum-well active regions separated and topped by
four (or more) linear gratings at different orientations.
Electrical contact is made to each grating and to a bottom contact
layer to measure four (or more) photocurrents. The device uses
interference among many light paths to encode in four
photocurrents four values that completely describe the
polarization state of the incident light at a given wavelength.  I
begin with the motivation for the new device.  Then, I report on
two computational models: the first includes perfectly conducting
gratings, and the second includes more realistic dielectric
gratings.  Next, I will describe the design and processing of a
two-layer proof-of-concept device.  Last, I will recount the
trials of the first round of fabrication.
\clearpage 
\end{abstract}

\tableofcontents

\listoffigures

\listoftables

\begin{glossary}{Longest  string}

   \item [$\alpha^\xi_m$]
       The wave vector component in the direction perpendicular to
       the grating groove in the medium $\xi=(a,b)$.
   \item [$\beta_m$]
        The $k_z$ component of the wave vector of mode $m$.  The
        real and imaginary parts of $\beta_m$ are always greater
        than or equal to zero.

   \item [BOE] Buffered oxide etch.  A solution of $HF$ buffered with $NH_3F$.

   \item [$C^{(\nu)}_{s,m,\pm}$]
        Coefficients in layer $\nu$ that represent the field in
        that layer.  The $s$ represents (fast, slow) or (TE, TM).
        The $m$ indicates the mode giving the x and y component of
        the wave vector.  The $\pm$ indicates an up-going or
        down-going mode.

   \item [$D(\nu,\nu+1)$]
        The transmission matrix relating the coefficients on one
        side of a slab interface to the other side of the
        interface.
   \item [${\mathcal{D}^{(\nu)}}$]  Dynamical matrix.  Used to find the transmission
   matrix.

   \item [$\vec{d}_j$]
      A set of discrete symmetries enumerated by $j$
      formed by translating the system by a displacement $\vec{d}_j$.

   \item[DOP]
      Degree of polarization.
   \item [Fast mode]
     The magnetic field parallel to the grating grooves vanishes.

   \item [$\gamma$]
       The wave vector component in the direction parallel to the
       grating grooves.

   \item [g] The coordinate parallel the grooves of a linear
   grating.

   \item [$\hat{g}$]
      A unit vector in the direction parallel to the grating
      groove.

   \item [h]  The coordinate perpendicular to the grooves of a
   linear grating.

   \item [$\Lambda_j$] Inherent polarization uncertainty.

   \item [$M$]
       The matrix that expresses the conditions for boundary
       condition matching for modes inside the dielectric
       gratings.

   \item [$\Phi$]  Optical phase difference between two paths in
   the device.

   \item [$PRM$]  Polarization response matrix: A matrix that relates the incident Stokes vector
   to the response vector.

   \item [QWIP]  Quantum-well infrared photoconductor.

   \item [$R$]  Response vector: A vector containing the response or light absorbed
   by each quantum-well stack in the device.

   \item[$S_0 \dots S_3$]
      The four stokes vectors.

   \item [${\mathcal{S}}$]
        The scattering matrix from the top reflective grating of
        the device.

   \item [Slow mode]
      The electric field parallel to the grating grooves vanishes.

   \item[TE mode]
       The electric field has no z-component.  Also known as `s'
       polarization.
   \item[TM mode]
       The magnetic field has no z-component.  Also known as `p'
       polarization.
   \item [$T(1,N)$]
        The transfer matrix.  Relates the fields at one end of a
        slab to the fields at the other extreme of the slab.

\end{glossary}

\mainmatter


\chapter{Polarimetry background}
\label{ChapPolBackground}

\section{Imaging polarimetry}

An imaging polarimeter captures an image with both the intensity
and some measure of the polarization state recorded for each
pixel.  A polarimetric image has more information than a simple
intensity image and improves remote sensing and automatic target
recognition~\cite{96Sadjadi01}.  Polarimetry can be used to
identify materials and to distinguish samples from a cluttered
background \cite{02Sposato01}.  Polarimetry has also shown promise
for mine detection~\cite{00Holloway01}.  On average, the pixels of
polarimetric images of man-made objects have a higher degree of
polarization (DOP) than the pixels of polarimetric images of
natural objects. This pattern could be useful for
spectro-polarimetric target detection with a target filling one
pixel or less.  Polarimetric data is often represented in terms of
Stokes parameters.

The four Stokes parameters\cite{75Born01}, which represent all the
polarization information, are: $S_0=I_0+I_{90}$,
$S_1=I_{0}-I_{90}$, $S_2=I_{45}-I_{135}$, and $S_3=I_R-I_L$ where
$I_X$ is the measured intensity of the light after passing through
a linear filter at an orientation of $X$ degrees, and $I_R$ and
$I_L$ are the measured intensities of the right or left circularly
polarized fraction of the light.

\begin{figure}
 \centerline{\includegraphics[width=5in]{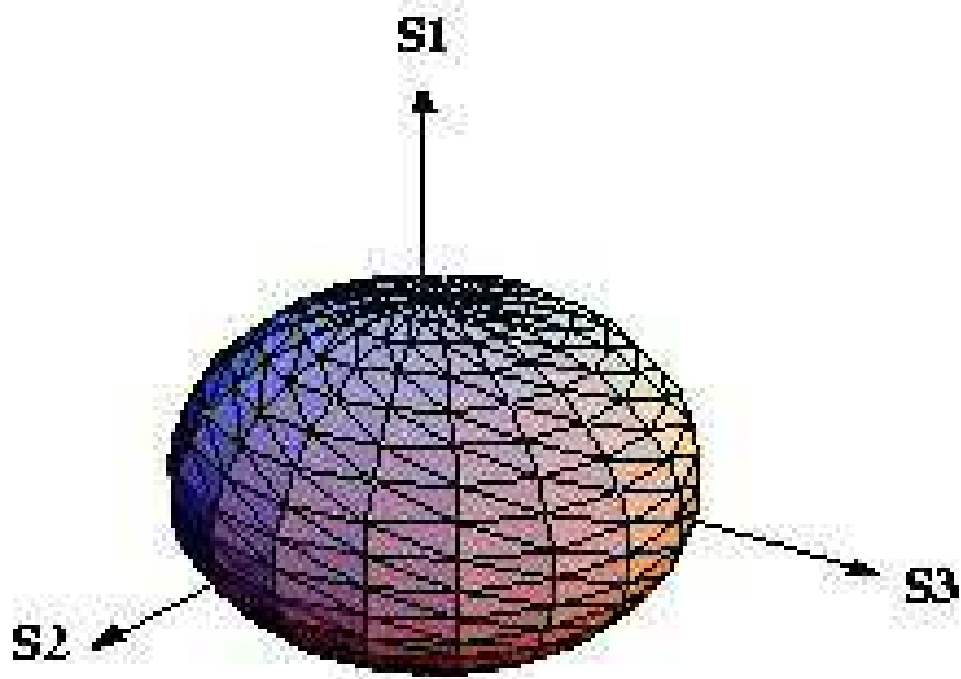}}
 \caption{\label{FigPoincareSphere} The Poincar$\acute{\rm{e}}$ sphere. A useful
 way of graphically representing the polarization of light in a
 3-dimensional vector space.  The polarization state is always
 some vector inside the sphere.  The radius of the sphere is the intensity
 of the light.}
\end{figure}

The Stokes parameters have several useful properties.  If you
place the four parameters into a four-dimensional vector (called a
Stokes vector), they form a linear vector space.  This means that
the Stokes vector of two beams of light added incoherently  is the
linear sum of the Stokes vector of each original beam.  The $S_0$
is the intensity of the light and is always positive $S_0>0$.  The
parameters will always satisfy the inequality ${S_1}^2 + {S_2}^2 +
{S_3}^2 \leq {S_0}^2$.  A useful way of representing the
polarization is by a vector in a ball, called a
Poincar$\acute{\rm{e}}$ sphere, shown in figure
\ref{FigPoincareSphere}. The radius of the ball is the intensity
of the light. Every polarization state is some vector inside the
ball or on the surface of the ball.  All points on the surface of
the ball are pure polarization states.  The degree of polarization
(DOP), is the ratio of the radius of the polarization state vector
to the radius of the ball.  In terms of Stokes parameters, the DOP
is given by $DOP=\sqrt{{S_1}^2+{S_2}^2+{S_3}^2}/S_0$.  From the
DOP and the above inequality, one can tell that $|S_1| \leq S_0$,
$|S_2| \leq S_0$, and $|S_3| \leq S_0$.

There are many approaches to imaging polarimetry instrument
design.  Some polarimeters only measure linear polarization and
intensity ($S_1$, $S_2$, and $S_0$).  Others will measure all four
Stokes parameters.  The choice of which combination of Stokes
parameters to measure depends on the application.  A
full-Stokes-vector polarimeter will measure all four Stokes
parameters at every pixel.

\section{Other polarimeter designs}

The current technology of imaging polarimeters cannot reliably
measure high-spatio-temporal polarization or
high-spectral-resolution polarization of a moving scene; the
camera reports huge errors at the boundaries of objects in the
scene or trades-off spatial and spectral resolution to achieve
faster measurements.

The errors at the boundaries of scene-objects occur because a
typical imaging polarimeter collects several different images of
the same object, with the light emanating from the object passing
through a different polarization filter in each image. The
collected images are linearly combined with positive and negative
weights to extract the polarization state at each pixel. To
perform the subtraction, one registers pixels from the different
images that represent the same point in the scene. However,
because the images are taken at different times, any motion in the
image will cause a registration error. Registration errors lead to
subtracting two values that should be of the same object but are
not of the same object. The difference will falsely report a
polarization in the image with errors in the range of 100\%
\cite{00Peterson01}.

U. S. Patent No. 5,045,701 by Goldstein and Chipman describes one
such device.  They use a rotating quarter-wave plate with a fixed
linear filter. With a single focal plane, this device takes
several images at different rotations, respectively, of the
polarizing filter.  Again, the collected images are linearly
combined with positive and negative weights to extract the
polarization state at each pixel. Again, one must register the
pixels from the different images that represent the same point in
the scene. As described before, any motion in the image will cause
a registration error. This precludes the use of this apparatus to
obtain polarimetric images of terrestrial objects from moving
platforms, such as airplanes or orbiting satellites, or images of
objects that are translating or rotating with respect to the
apparatus.

Beekman and Van~Anda used neighboring pixels to detect different
polarizations; again the pixels are misregistered by one pixel
width \cite{01Beekman01}.  This approach uses quantum well
infrared photodetectors with linear gratings. Quantum wells can
only detect the component of light with the electric field
perpendicular to the growth direction \cite{93Levine01,98Liu01}.
With linear gratings, each pixel can only detect the component of
the incident light with the electric field perpendicular the
grooves of the grating.  On a single focal plane, Beekman and
Van~Anda made neighboring pixels sensitive to vertical, horizontal
or diagonal polarizations by using vertical, horizontal or
diagonal gratings. Again, one linearly combines the neighboring
pixels sensitive to the different polarizations with positive and
negative weights. Because the pixels being combined image
spatially neighboring points in the scene, sharp edges or bright
points will register as erroneous polarization.

Another approach uses four separate cameras with a different
polarization filter on each camera.  The four cameras take
simultaneous images of the same scene. Again, the appropriate
images are linearly combined with positive and negative weights
to extract the four Stokes parameters; however, parallax and
camera misalignment will introduce registration errors into the
derived image.  Polarization error follows due to registration
errors.

In a brand new type of spectral-polarimeter, pixel registration is
achieved but spectral resolution is sacrificed
\cite{99Iannarilli}. In this case, the pixels reading nearby
groups of spectral lines contain the polarization data.  Due to
this source of error, sharp spectral features will induce
polarization errors. This example is called a polarimetric
spectral intensity modulation spectropolarimeter, and is described
in U. S. Patent No. 6,490,043 by Kebabian.  This device measures
the polarization of a single point in a scene by modulating the
spectrum of the light with the polarization of the light and then
measuring the spectrum of the light on a row of pixels.  In order
to find the polarization one must compare the modulated spectrum
to the true spectrum. Because the true spectrum is not known,
approximations must be made that necessarily sacrifice
polarimetric and spectral accuracy and precision in favor of pixel
registration.   Their approach of encoding the polarization in the
spectrum is very similar to approaches described at the Japanese
SICE conference by Oka~\cite{02Oka01} and Locke~\cite{01Locke01}.
Another polarimeter design that also records the polarization as a
spatial pattern on the focal plane array is explained by Van
Delden \cite{03VanDelden01}.

All current polarimeters are lacking in one of three areas. First
is registration errors: the error is maximum at the boundaries of
objects in a scene.  The second difficulty is sacrificing spectral
resolution: sharp spectral features still introduce errors.  The
third is a sacrifice in spatial resolution:  one must trade-off
the weight and volume of imaging optics, which normally limit
resolution, with the weight and volume of beam splitters needed to
divide the image onto several focal-plane arrays.


\chapter{Introduction to a polarimeter-in-a-pixel}
\label{ChapConceptOfOperation}

The topic of this thesis is a device that should detect the
full-polarization at each pixel, and at each frame, and at each
wavelength.  The device was first introduced in reference
\cite{02Serna01}, and was also reported on in references
\cite{03Serna02,03Serna01}.  In this chapter, I describe the
device.

\section{Device description}

\begin{figure}
 \centerline{\includegraphics[width=5in]{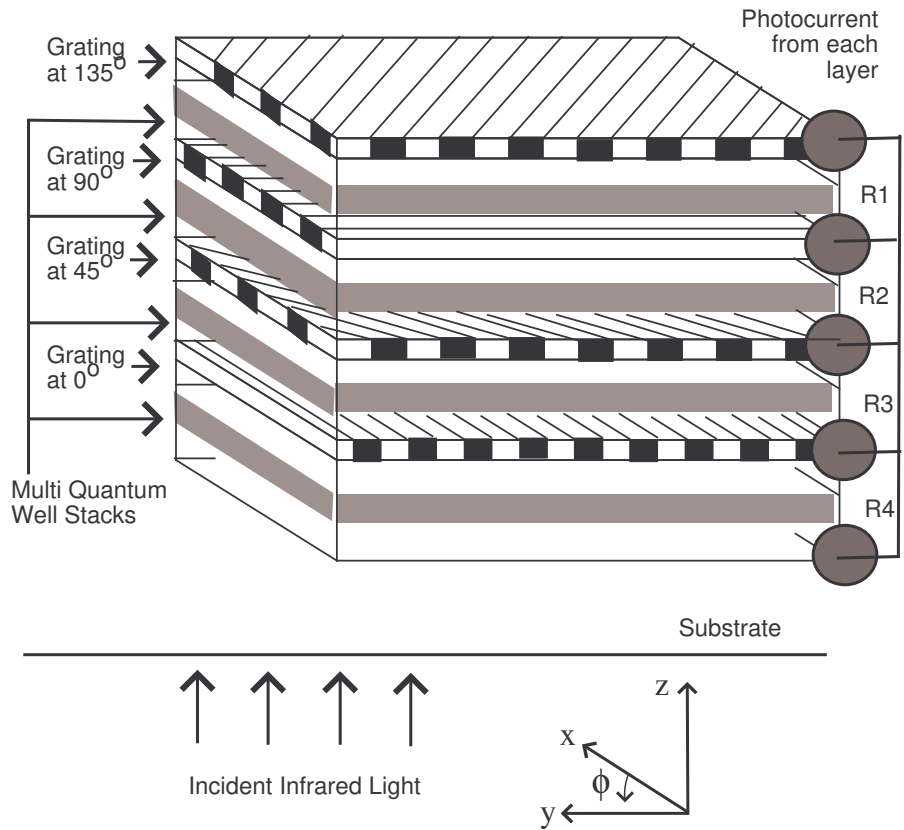}}
 \caption{\label{FigPolarimeter}\label{FigDevice} The polarimeter-in-a-pixel.
 A stack of multi-quantum-well photoconductors and gratings at
 different orientations.  The photocurrent from each layer is
 read individually.}
\end{figure}
In the device, shown in figure \ref{FigDevice}, quantum-well
stacks are used in combination with linear gratings to determine
the degree of polarization of incident light in terms of Stokes
parameters. Interference from multiple reflections, diffractions
and transmissions of the light propagating from and through the
linear gratings modulates the absorption of the $\pm$1-diffracted
orders at each quantum-well stack. The quantum-well stacks, do not
absorb light having an electric field component in a plane
parallel to the quantum well stacks \cite{93Levine01,98Liu01}.
The non-absorbed propagating light is reflected, diffracted and
transmitted at each grating as a function of its polarization.
Interference translates the incident polarization into the amount
of polarized light having an electric field with a component in
the z-direction. This z-polarized component of the light is
absorbed by the quantum wells.

Each quantum-well stack is included in a separate circuit having a
voltage bias and a current meter. The voltage bias across each
circuit is individually adjusted, and the photocurrent in each
circuit, as measured by the respective current meter, is
proportional to the flux of light absorbed by the respective
quantum-well stack.  The circuit diagram is similar to that of
multi-color QWIP devices \cite{01Goldberg01}.

The four photocurrents are thus a measure of the polarization of
the incident light by measuring what is ultimately absorbed by
each of the four quantum well stacks.  More particularly, the four
currents are linearly mapped to the four Stokes parameters, which,
in turn, represent the polarization of the incident light.

\section{Principle of operation}

To explain the principle underlying the polarization detection,
consider a current-technology quantum-well photodetector including
single quantum-well stack with a single linear grating and
consider a simplified embodiment of the present invention with
just two quantum-well stacks and two linear gratings. Using the
single layer of gratings, this section will show the relevant
physics of the quantum-well polarization selection rule for
absorption, and the relevant physics of diffraction from a single
grating. Using the two-layer structure, this section will show how
the incident photons are split and then interfere to gain
additional polarization sensitivity.

\subsection{Single-layer device}

\begin{figure}
 \centerline{(a)\includegraphics[width=3in]{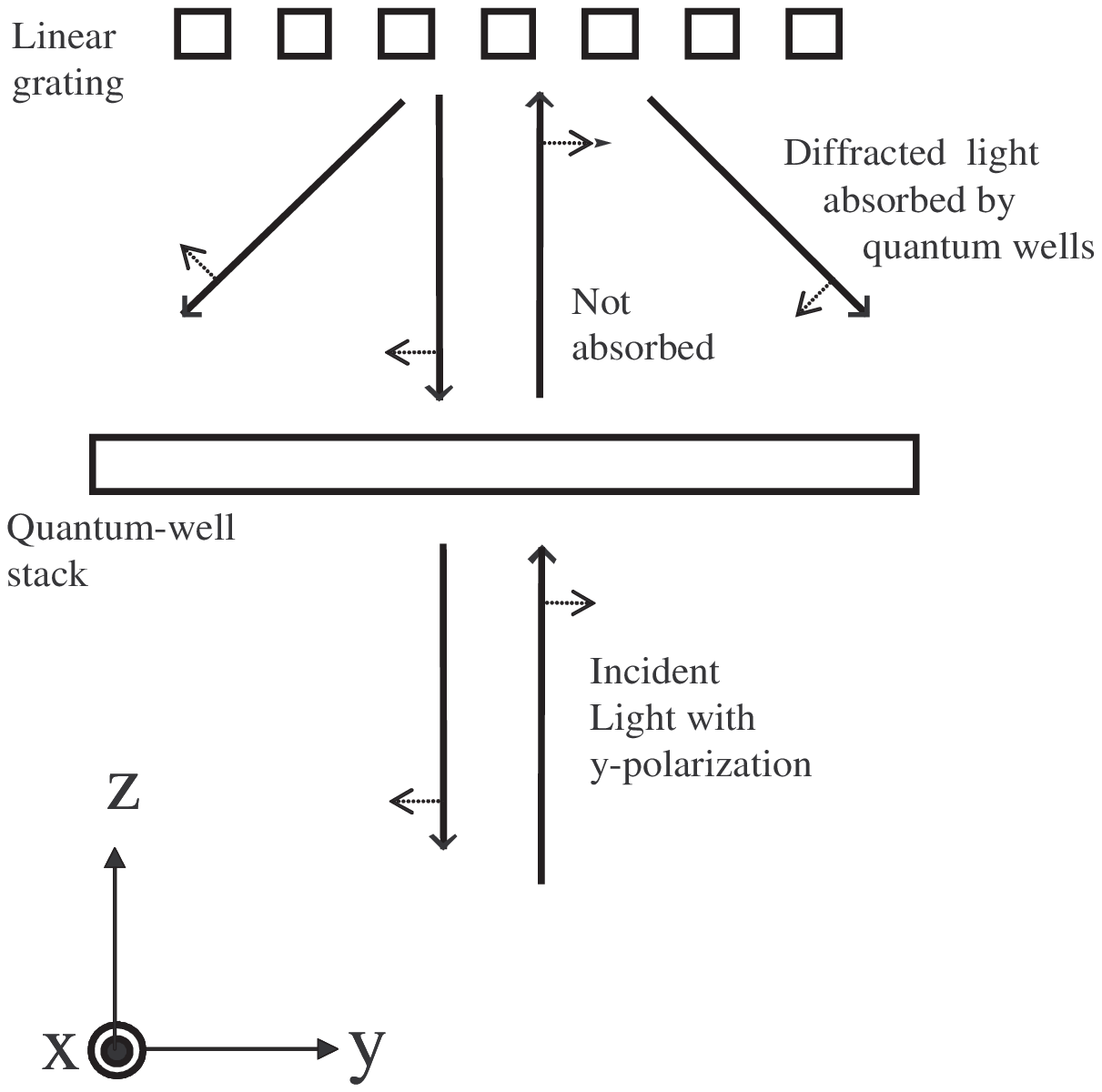}(b)\includegraphics[width=3in]{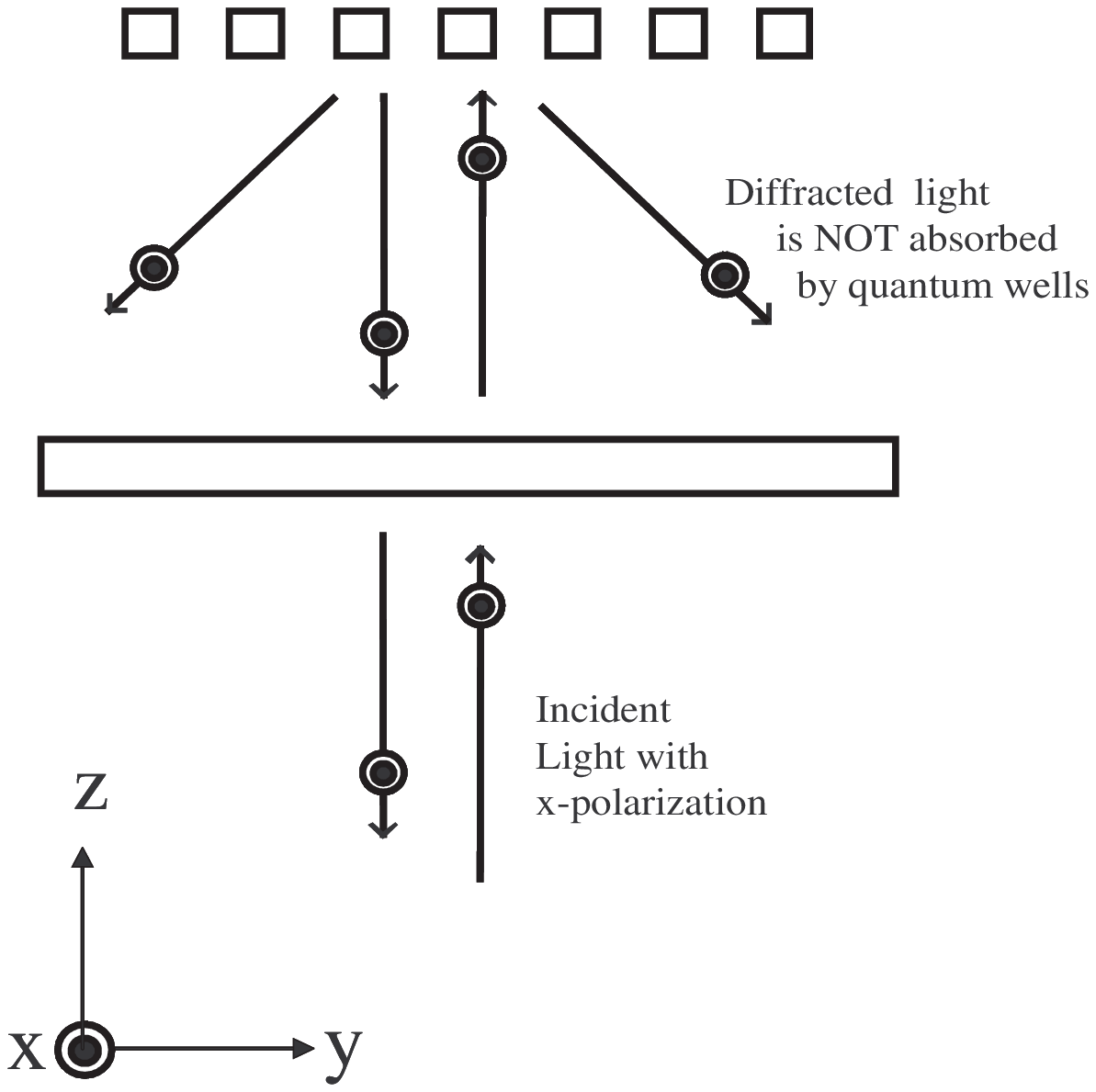}}
 \caption{\label{FigSingleLayerQWIP} A traditional quantum-well
 infrared photoconductor (QWIP) with a linear grating along the x-axis.
 The QWIP will only absorb y-polarized incident light due to the nature of
 the diffraction gratings and the polarization selection rule of the quantum wells.}
\end{figure}

The quantum wells, and thus the quantum-well stacks, only absorb
light with an electric field in the z-direction
\cite{93Levine01,98Liu01}. Figures \ref{FigSingleLayerQWIP} (a)
and (b) show two different polarizations of light propagating
through a current-technology quantum-well infrared detector(QWIP).
The QWIP is equipped with a linear grating in the x-direction (out
of the page).  This subsection will show that only the y-polarized
light is absorbed.

Because the electric field is always perpendicular to the
direction of travel, light propagating in the z-direction is never
absorbed.  Due to this property of light, and the polarization
selection rule, the incident beam and the reflected beam are not
absorbed -- only the diffracted light has a chance of being
absorbed.

Figure \ref{FigSingleLayerQWIP} (a) shows how the diffraction
grating enables light to be absorbed by the quantum wells.  The
incident light is not absorbed on the first pass through the
quantum well material.  The reflected light is also not absorbed.
The diffracted light has the electric field shifted into a
component parallel to the z-axis.  This light is absorbed.

A linear grating preserves the fast polarization (electric field
parallel to the grating grooves) of light incident from the plane
perpendicular to the grooves of the grating. Thus, the x-polarized
light will be diffracted into x-polarized diffracted orders. As
shown in figure \ref{FigSingleLayerQWIP} (b), the diffracted
x-polarized light is still perpendicular to the z-axis and is not
absorbed by the quantum-wells.

In this way, a simple QWIP with a linear grating can only detect
the y-polarized component of the incident light.  However, the
single pixel just described will give the same response for a
fixed intensity of 45-degree polarized light, 135-degree polarized
light, right-circularly polarized light, left-circularly polarized
light, and unpolarized light.

\subsection{Two-layer test structure}

\begin{figure}
 \centerline{\includegraphics[width=5in]{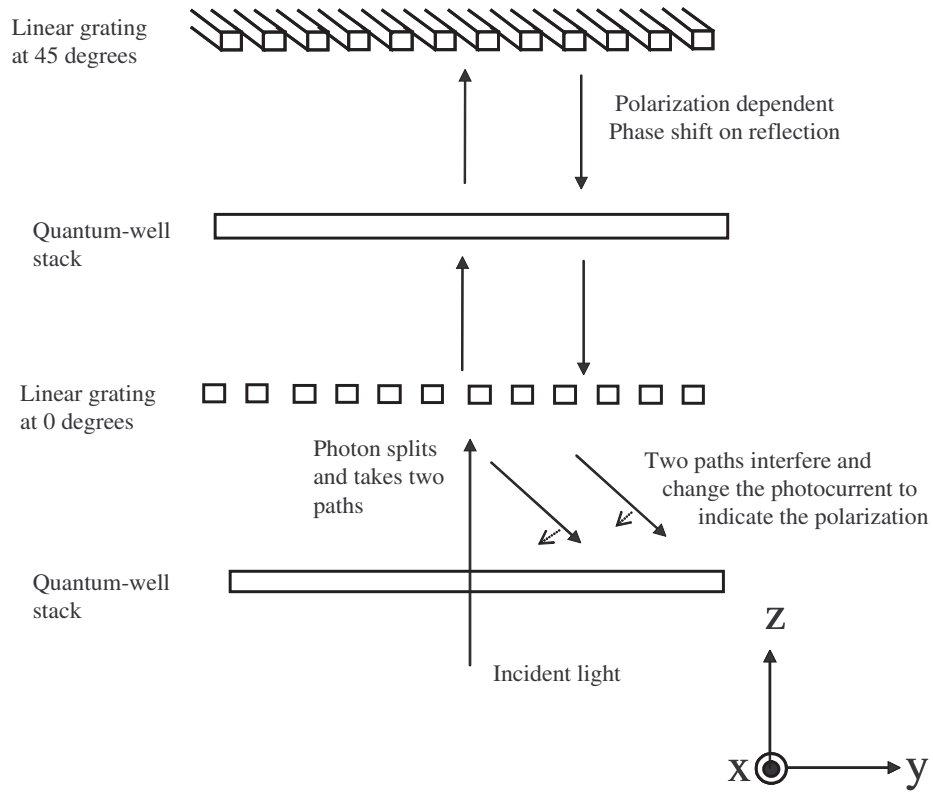}}
 \caption{\label{FigTwoLayerRayTrace}  The incident photons are split into
 different paths that interfere.  This interference is the origin of
 the polarization sensitivity of the proposed device. }
\end{figure}

To understand how my new device works, this section follows two
paths of light through a simplified two-layer embodiment of the
invention.  To show an improvement over the old technology, I need
to show how this new invention can distinguish 45-degree from
135-degree polarization in a single pixel.

The non-absorbed z-propagating light is reflected, diffracted and
transmitted at each grating as a function of its polarization. One
key element is that linear gratings also have
polarization-dependent physics \cite{80Petit01,88Kok01}. Grating
diffraction efficiency is different for fast polarization (no
magnetic field parallel to the grating's grooves) and slow
polarization (no electric field parallel to the grating's
grooves).

For example, consider light with a 2.5 $\mu m$ wavelength normally
incident on a perfectly conducting grating with a 3 $\mu m$
period, 1.5 $\mu m$ grooves, and  0.75 $\mu m$ depth.  The fast
polarization will be 14$\%$ reflected (into the  0$^{th}$ order)
and 43$\%$ diffracted into the $\pm 1$ orders, and the
slow-polarization will be 30$\%$ reflected (into 0$^{th}$ order)
and 35$\%$ diffracted into the $\pm 1$ orders. In addition,
polarized light incident on a linear grating will have a relative
phase shift between the diffracted fast and slow polarization
components~\cite{88Kok01}.

In the case of a single grating, for a fixed intensity of light,
we could distinguish x-polarized from y-polarized, but we could
not distinguish 45-degree-polarized from 135-degree-polarized. I
will now describe how interference from the second linear grating
at an angle of 45 degrees will modulate the photocurrent measured
across the quantum wells before the first grating.

Each incident photon is split at each grating and takes several
paths through the device.  Figure \ref{FigTwoLayerRayTrace} shows
an example of two paths that light can take that create
polarization dependent interference and therefore lead to
polarization dependent photocurrent.  Recall, the actual device
has four or more layers and many interfering paths.  I will just
trace through two simple paths.

For 45-degree-polarized incident light, the first path consists of
a diffraction of light from the grating along the x-axis.  I
define that when the incident light reaches the first grating, the
phase is 180 degrees.   After diffraction, the component of the
reflected +1 order that is polarized in the y-z plane has a phase
of 105 degrees.

For 45-degree-polarized incident light, the second path consists
of the transmission of the incident light through first grating.
Transmission through this grating causes the light to become
somewhat elliptically polarized with a major axis along the
45-degree orientation.  The component of the light polarized in
the y-z plane has a phase of -90º.  The propagation of light from
first grating to grating second grating advances the phase by 400
degrees. At the second grating, the light reflects backward, and
the light component in the y-z plane now has a phase of 415
degrees. The magnitude and the phase shift in the light after
reflection from a linear grating is highly dependent on the
polarization of the light with respect to the grating. The light
advances another 400 degrees in phase while propagating back
towards the first grating. After the light forward diffracts from
this first grating, the phase becomes 915 degrees, which is
equivalent to 105 degrees + 90 degrees.  We compare with 105
degrees because this was the phase of the light that took the
first path, \emph{i.e.} was diffracted and reflected to the +1
order.

Since the phases polarized in the y-z plane from these two paths
do not cancel out, the quantum-well stack will absorb the light
and generate a photocurrent. The absorption by quantum-well stack
is proportional to the vector sum of the light from these two
paths. The two paths have a phase difference of 90º, and therefore
the two paths do not interfere constructively or destructively.
This leads to a measurable amount of photocurrent across quantum
well stack.

Next consider 135-degree polarized incident light. Again, I define
that when the incident light reaches the first grating, the phase
is 180 degrees. Since the light is normally incident, the
quantum-well stack cannot absorb the light as it first enters the
device .  Again, interference exists between two possible paths
that the photon can take.  The interference is found by following
a photon through two paths, which are in principle the same
photon.

For 135-degree-polarized incident light, the first path consists
of a diffraction from the first grating to the reflected and
diffracted +1 order. The component of the diffracted light
polarized in the y-z plane will have a phase of 105 degrees.

For 135-degree-polarized incident light, the second path consists
of the transmission of the incident light through the first
grating. After transmission through this first grating, the light
is slightly elliptically polarized with a major axis along the
135-degree orientation.  The component of the light polarized in
the y-z plane has a phase of -90 degrees.  The propagation of the
light from the first grating up to the second grating advances the
phase by 400 degrees.  After reflection from the second grating,
the light as a phase in the y-z plane of 505 degrees.  This is the
step that differentiates between incident light polarized along
45-degree and 135-degree axes.  The reflection from the second
grating gives a very different phase shift when the light is
polarized with versus perpendicular to the grating grooves. Next,
the light advances another 400 degrees in phase while propagating
back towards the first grating.  After the light forward diffracts
from this first grating it is now in the same mode as the light
that originally reflected and diffracted into the +1 order.  The
phase in the y-z plane is now 1005 degrees, which is equivalent to
105 degrees + 180 degrees.

The absorption by quantum-well stack is proportional to the vector
sum of light from both paths.  The two paths have a phase
difference of 180 degrees, and therefore the two paths interfere
destructively.  This leads to a negligible amount of photocurrent
across quantum-well stack.  Therefore, the photocurrent across
this first quantum-well stack changes for 45-degree polarized
incident light compared to 135-degree polarized incident light.

In this example, I have traced through two paths from a two-layer
device.  To predict the actual photocurrents of the actual device,
I would need to trace through thousands of paths in a structure
with four or more layers.  This process is done through a computer
model described in the next chapter.


\chapter{Modeling the complete four-layer polarimeter-in-a-pixel}
\label{ChapModeling}

In order to validate the principle-of-operation described in the
previous chapter, I developed a working electromagnetic model of
the device using transfer-matrix techniques described in Andersson
\cite{91Andersson01,92Andersson01}  and Wendler
\cite{99Wendler01}.  The code was written in GNU C++ on
windows-based computers and is available upon request.

\begin{figure}
 \centerline{
 \includegraphics[width=3.4in]{./GrayScaleDiagram.eps} }
 \centerline{
 \includegraphics[width=6.7in]{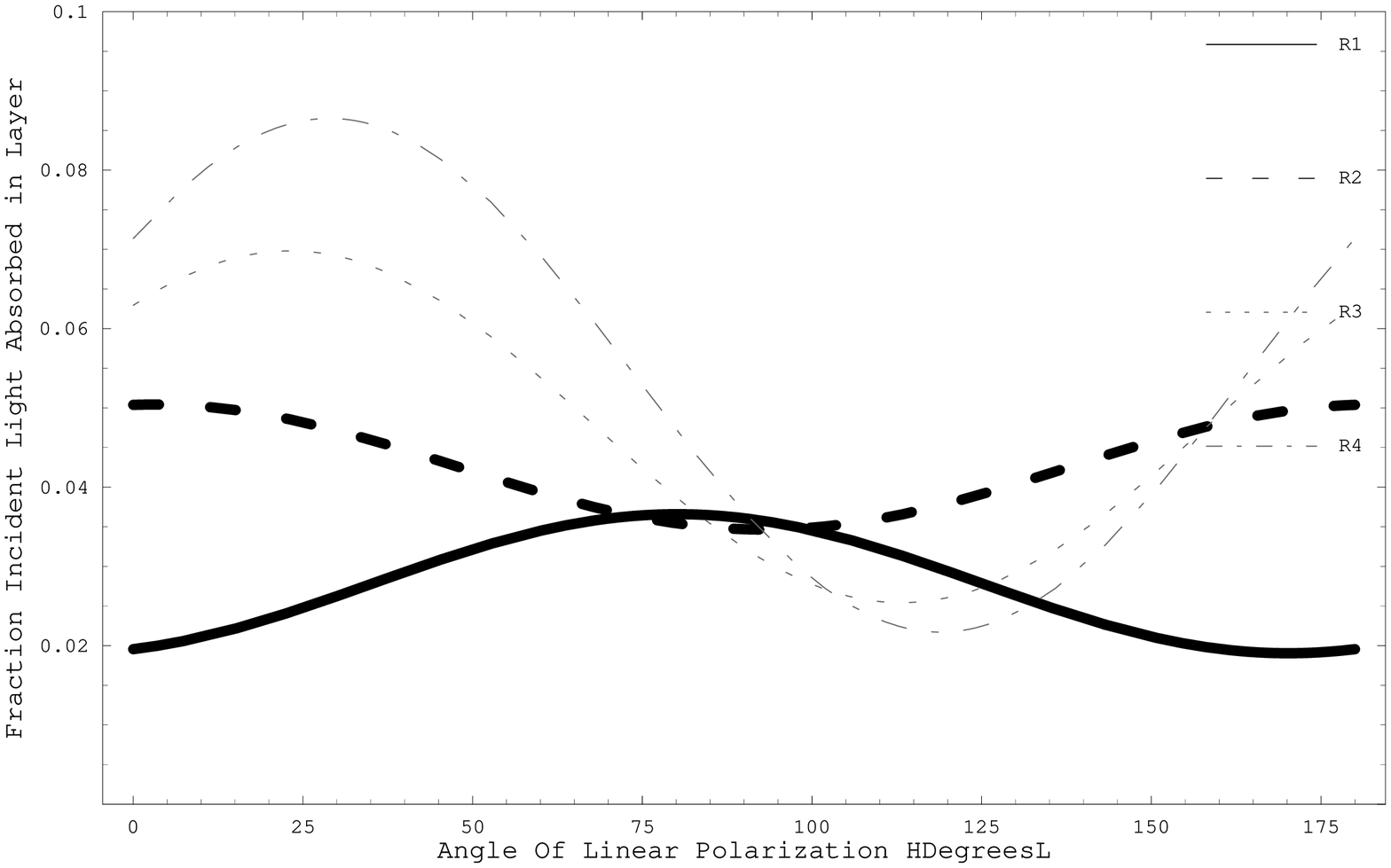}}
 \caption{\label{FigLinResponce}
 An example of the response from the four layers of the pixel-polarimeter
 as we rotate the polarization angle of the incident linearly-polarized light.}
\end{figure}

\begin{figure}
 \centerline{
 \includegraphics[width=3.4in]{./GrayScaleDiagram.eps}}
 \centerline{
 \includegraphics[width=6.7in]{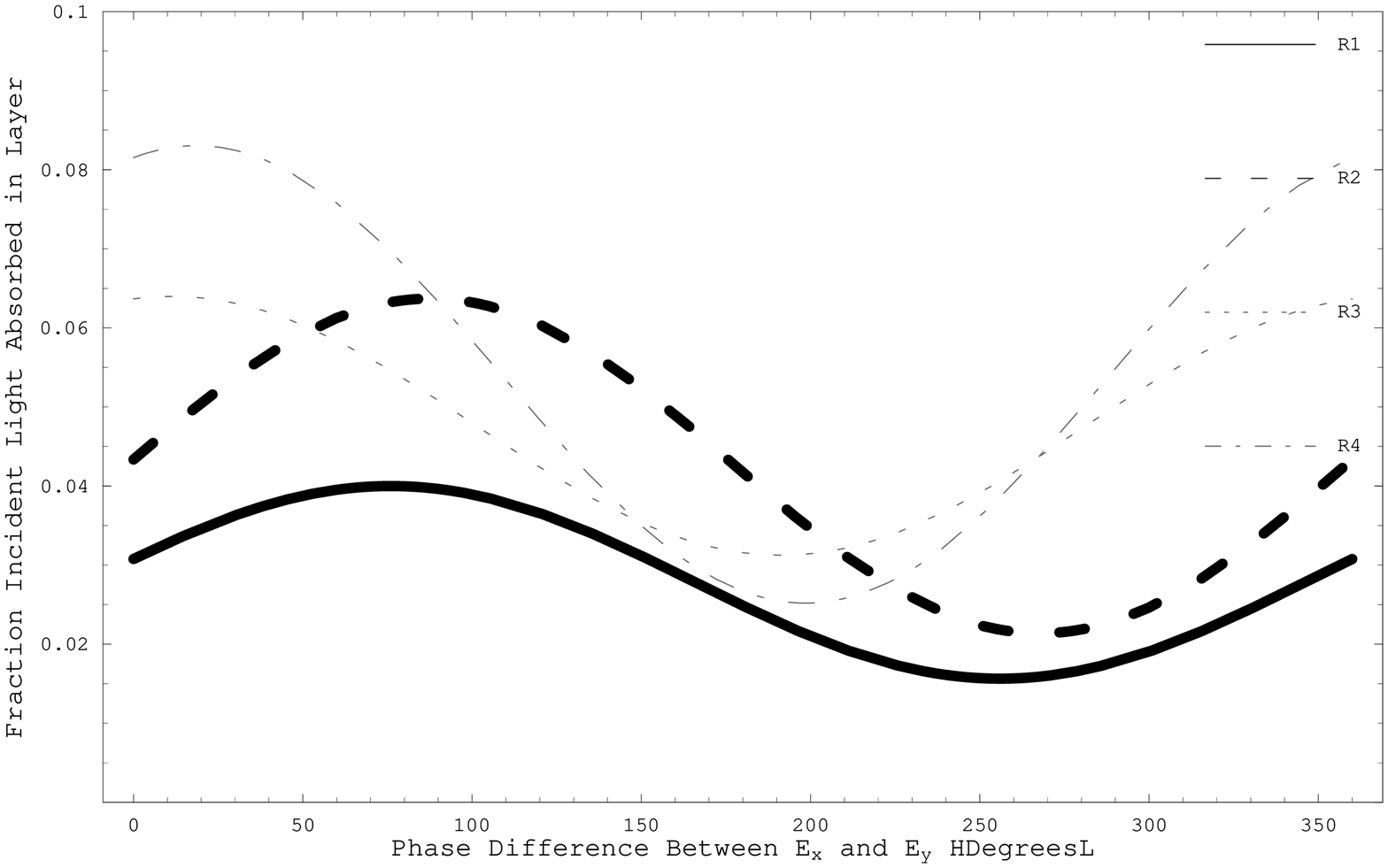}
 }
 \caption{\label{FigCircResponce} An example of the response from
 the four layers of the pixel-polarimeter as
 we vary the phase lag between an x-polarized and y-polarized
 plane wave.}
\end{figure}

Figures \ref{FigLinResponce} and \ref{FigCircResponce} show some
results of this working electromagnetic model of the proposed
device. On the graph's ordinate, the fraction of incident light
absorbed in the four quantum-well stacks respectively for 9.5
micron incident light.  The abscissa of figure
\ref{FigLinResponce} shows the angle of linearly polarized
incident light. The abscissa of figure \ref{FigCircResponce} shows
the phase lag between 0-degree polarized incident light and
90-degree polarized incident light. The phase lag smoothly changes
the polarization from 45-degree linear to right circular to
135-degree linear to left circular.  The graph demonstrates that
the relative photocurrents from the four quantum-well stacks
provide a means to measure the polarization of incident light.

\section{The transfer-matrix method}


The model that generated figures \ref{FigLinResponce} and
\ref{FigCircResponce} is based on a transfer-matrix method. A
transfer-matrix approach breaks the problem down into slabs. The
modes in each slab are found separately.  Boundary conditions at
the interface between the slabs relate the electromagnetic state
in one slab to the neighboring slabs. Each slab must be uniform
along the z-axis. The $N$ slabs are designated by the index $\nu$:
\begin{eqnarray}
\nu=N &  &  h_{N-1} < z \\
\vdots & & \vdots \\
\nu=3 & & h_2 < z < h_3 \\
\nu=2 & & h_1 < z < h_2  \\
\nu=1 &  &  \ \ \ \ z < h_1,
\end{eqnarray}
where $h_\nu$ are the z positions of the interfaces between each
slab.

The fields in each slab, $F^{(\nu)}$ (either an electric or
magnetic field) are represented by a sum of coefficients
multiplied by mode functions,
\begin{equation}
F^{(\nu)}_s(x,y,z) = \sum_{m,\pm} C^{(\nu)}_{s,m,\pm}
\Psi^{(\nu)}_{s,m,\pm}(x,y,z),
\end{equation}
where $C^{(\nu)}$ are the coefficients and $\Psi^{(\nu)}$ are the
mode functions. The index $m$ represents the x, y component of the
wave vector for that particular mode. Up-going modes are denoted
($+$) where $k_z
>0$, and down-going modes (-) where $k_z < 0$.  The index $s$
indicates which of two states one is in for each wave vector. The
two states could be TE and TM or fast and slow.  For TE modes, the
electric field has no z-component.  For TM modes, the magnetic
field has no z-component. For fast modes, the magnetic field
parallel to the grating grooves vanishes.  For slow modes, the
electric field parallel to the grating grooves vanishes.

  The field $F^{(\nu)}_s$ is chosen so that
the 6 components of the electromagnetic field are implicitly
stored in at most 2 components.  For example, the electric field
in the direction $\hat{k} \times \hat{z}$ for TE modes and the
magnetic field in the direction $\hat{k} \times \hat{z}$ for TM
modes. Given these two components and the wave vector $\vec{k_m}$
for the particular mode, one can find all 6 components of the
electric and magnetic vector fields.

The coefficients $C^{(\nu)}_{s,m,\pm}$ in each slab are collected
into a vector that succinctly represents the fields in that slab.

Because the tangental fields are continuous across the boundary of
each slab, we can easily find the coefficients that represent the
field in slab $\nu$ if we know the coefficients in slab $\nu+1$.
The matrix that relates these two sets of coefficients is called a
transmission matrix.  The transmission matrix $D(\nu,\nu+1)$
converts the coefficients in slab $\nu+1$ at $z=h_\nu$ to the
coefficients in slab $\nu$ at $z=h_\nu$:
\begin{equation}
C^{(\nu)}= D(\nu, \nu+1)\, C^{(\nu+1)}.
\end{equation}
Within a slab $\nu$, one can also relate the fields at
$z=h_{\nu+1}$ to the fields at $z=h_{\nu}$ by a diagonal
propagation matrix,
\begin{equation}
 \left( \matrix{ C^{(\nu)}_{s,m,+}(h_\nu) \cr
                 C^{(\nu)}_{s,m,-}(h_\nu)} \right)
 = \left( \matrix{ \exp(-i\,{k_z}^{(\nu)}_{s,m} \Delta_h) & 0 \cr
     0 & \exp(+ i \,{k_z}^{(\nu)}_{s,m} \Delta_h) }\right)
 \left( \matrix{ C^{(\nu)}_{s,m,+}(h_{\nu+1}) \cr
                 C^{(\nu)}_{s,m,-}(h_{\nu+1})} \right)
\end{equation}
where $\Delta_h=h_{\nu+1}-h_{\nu}$ and both the real and imaginary
parts of $k_z$ are taken to be positive.

Using the same parameters as reference \cite{92Andersson01}, the
quantum wells were represented with a Drude model where
$\epsilon_{zz}$ was imaginary and represented absorption.

By stringing together the transmission matrices and the
propagation matrices, we can relate the fields at one extreme of
the stack to the fields at the other end of the stack through the
transfer matrix $T(1,N)$:
\begin{equation}
T(1,N)=D(0,1)\prod_{\nu=1}^{N-1}\, P(h_\nu,h_{\nu+1})\,
D(\nu,\nu+1).
\end{equation}

To use the transfer matrix, we need to know boundary conditions at
one or both ends of the stack.  In our case, the incident light is
given. Therefore we know the coefficients of the up-going modes in
layer $\nu=1$.  This forms one boundary condition. Also, the top
grating is completely reflecting so the upward and downward
traveling light at layer $\nu=N$ are related via a scattering
matrix,
\begin{equation}
   C^{(\nu=N)}_{s,m,-} = {\mathcal{S}}^{s',m'}_{s,m} \
   C^{(\nu=N)}_{s',m',+}.
   \label{EqScatteringMatrix}
\end{equation}

If I break up the transfer matrix T(1,N) into four matrices
relating the up-going and down-going coefficients,
\begin{equation}
 \left( \matrix{ C^{(\nu=1)}_{+}(h_1) \cr
                 C^{(\nu=1)}_{-}(h_1)} \right)
 = \left( \matrix{ T(1,N)_{++} & T(1,N)_{+-} \cr
                   T(1,N)_{-+} & T(1,N)_{--} }\right)
 \left( \matrix{ C^{(\nu=N)}_{+}(h_{N-1}) \cr
                 C^{(\nu=N)}_{-}(h_{N-1})} \right),
\end{equation}
where the $s,m$ indexes are suppressed, then, by using the
boundary conditions, I find the fields at layer $\nu=N$, in terms
of the incident coefficients $C^{\nu=1}_+(h_1)$ are given by
\begin{equation}
C^{(\nu=N)}_{+}(h_{N-1})= \left[\, T(1,N)_{++}\ + \ T(1,N)_{+-}\
{\mathcal{S}} \ \right]^{-1}\ C^{(\nu=1)}_+(h_1).
\end{equation}
Using the scattering matrix in equation \ref{EqScatteringMatrix},
I find the down-going modes at layer $\nu=N$ from the up-going
modes.  Using the transfer matrix, these modes can be propagated
to any layer in the structure.

The transfer matrix takes into account infinite reflections and
interference.  This is the main advantage of the method.  However,
the solutions of thick structures, where $h_N - h_1 >>
\frac{1}{\rm{max}_m ( Im\{ k_{z\,s,m} \} )}$, involve inverses
that become unstable. Therefore, the transfer-matrix method
degrades as the structure being modeled becomes thicker and higher
order modes are considered.

\section{Transmission matrices for the quantum-well slabs}

The first type of transmission matrix we need is for the
quantum-well slabs. The interface is assumed smooth so the modes
are not mixed; we need only consider reflections and
transmissions. The transmission matrix is given by
 \begin{equation}
 \left( \matrix{ C^{(\nu)}_{s,m,+} \cr
                 C^{(\nu)}_{s,m,-} } \right)
 =
 \frac{1}{t_{s}(\nu,\nu+1)}\left( \matrix{1 & r_{s}(\nu,\nu+1) \cr
                                            r_{s}(\nu,\nu+1) & 1} \right)
 \left( \matrix{ C^{(\nu+1)}_{s,m,+} \cr
                 C^{(\nu+1)}_{s,m,-} } \right),
                 \end{equation}
where $r_s$ and $t_s$ are the Fresnel reflection and transmission
cofficients respectively and $s$ is an index which indicates if we
want the TE or TM case.

In the case of anisotropic media, the Fresnel coefficients need to
be more carefully defined. The coefficients are normally defined
in terms of a plane wave with the electric field perpendicular to
the plane of incidence (TE), or a plane wave with the electric
field parallel to the plane of incidence (TM). Also, the
traditional definition of the angle of incidence and angle of
transmission is insufficient in uniaxial media because the
dispersion relation relating $k_x$, $k_y$, and $k_z$ is different
for $TE$ and $TM$ polarization.

To clarify, I present the dispersion relation for a uniaxial
non-istropic material in terms of the wave vector components. The
dielectric constant is the same for the $\hat x$ and $\hat y$
directions, denoted $\epsilon_\parallel$ and different for the
$\hat z$ direction, denoted $\epsilon_{zz}$.  Due to the
continuous translational symmetry parallel to the flat interfaces
of the layers, the vector $k_{\parallel}$ is preserved across the
interfaces. For $TE$ polarization we have
 \begin{equation}
 k_{\parallel}^2+k_z^2=\epsilon_\parallel \frac{\omega^2}{c^2},
 \end{equation}
and for $p$ polarization we have
 \begin{equation}
 k_{\parallel}^2+\frac{\epsilon_{zz}}{\epsilon_\parallel} k_z^2=\epsilon_{zz}
 \frac{\omega^2}{c^2}.
 \end{equation}

For future convenience, the real part of the electric field for TE
modes is in the direction $\hat{k} \times \hat{z}$.  The real part
of the magnetic field for TM modes is in the direction $\hat{k}
\times \hat{z}$.

The Fresenel reflection coefficient for $TE$ polarization from
layer $\nu$ to layer $\nu+1$ is,
 \begin{equation}
 r_E(\nu,\nu+1)= \frac{\hat{n} \cdot (\vec k_{\nu} - \frac{\mu_\nu}{\mu_{\nu+1}} \vec k_{\nu+1})}
 {\hat{n} \cdot (\vec k_{\nu} + \frac{\mu_{\nu}}{\mu_{\nu+1}}\vec k_{\nu+1})}
 \end{equation}
and the transmission coefficient,
 \begin{equation}
 t_E(\nu,\nu+1)= \frac{2 \hat{n} \cdot \vec k_\nu}{\hat{n} \cdot (\vec k_\nu + \frac{\mu_\nu}{\mu_{\nu+1}}
 \vec k_{\nu+1})}
 \end{equation}
where $k_\nu$ is the wave vector in the medium of the $\nu$th
layer, likewise $\mu_1$ and $\mu_2$ are the magnetic
susceptibilities of the incident and transmitted material
respectively, and $\hat{n}$ is the normal vector to the surface
which in our case is always $\hat{z}$.

For $TM$ polarization, the transmission and reflection
coefficients are
 \begin{equation}
 t_M(\nu,\nu+1)= \frac{2 {\epsilon_{\nu+1}}_{xx} \hat{n} \cdot \vec k_\nu}
 {\hat{n} \cdot ({\epsilon_{\nu+1}}_{xx} \vec k_\nu +  {\epsilon_\nu}_{xx} \vec k_{\nu+1})}
 \end{equation}
and
 \begin{equation}
  r_M(\nu,\nu+1)= \frac{\hat{n} \cdot ({\epsilon_{\nu+1}}_{xx} \vec k_\nu -  {\epsilon_\nu}_{xx} \vec k_{\nu+1})}
 {\hat{n} \cdot ({\epsilon_{\nu+1}}_{xx} \vec k_\nu +  {\epsilon_\nu}_{xx} \vec
 k_{\nu+1})}.
 \end{equation}

\section{Modes in the bulk structure}

I am solving a problem involving gratings. A tutorial introduction
to the methods of solving diffraction grating problems is given by
Petit in Ref.~\cite{80Petit01}.   Grating periodicity places
limits on the modes that I can consider in the bulk.  I
approximate the periodic grating to have an infinite extent. For
electromagnetic waves of frequency $\omega$ in the presence of a
linear grating, the fields must satisfies the quasi periodicity
condition,
\begin{equation}
f(y)=f(y+d) exp(i\, k_o\, y),
\end{equation}
where $k_o$ is the component of the incident wave vector
perpendicular to the grooves of the grating. The fields can then
be expressed as a sum of a discrete set of modes,
\begin{equation}
F(y,z)=\sum_{m,\pm} C_{m,\pm} \exp( i \alpha_m y \pm i \beta_m z)
\label{modalexpansion}
\end{equation}
where $C_m$ are unknown complex constants that span the solution
space and are determined by matching the boundary conditions,
\begin{equation}
\alpha_m = m \Delta k_y + k_o,
\end{equation}
\begin{equation}
\beta_m = \sqrt{k^2 - \alpha_m^2},
\end{equation}
\begin{equation}
k=\frac{2 \pi\ c\ n(\omega)}{ \omega},
\end{equation}
and where $n(\omega)$ is the index of refraction of the material
and $\Delta k_y = \frac{ 2 \pi}{d}$.  When $\alpha_m^2 > k^2$, I
take $\beta_m$ to be the positive imaginary root.  The terms with
$\beta_m \in \rm{Im}$ are referred to as evanescent modes.  Terms
with $\beta_m \in \rm{Re}$ are called propagating modes.

\begin{figure}
 \centerline{\includegraphics[width=3.5in]{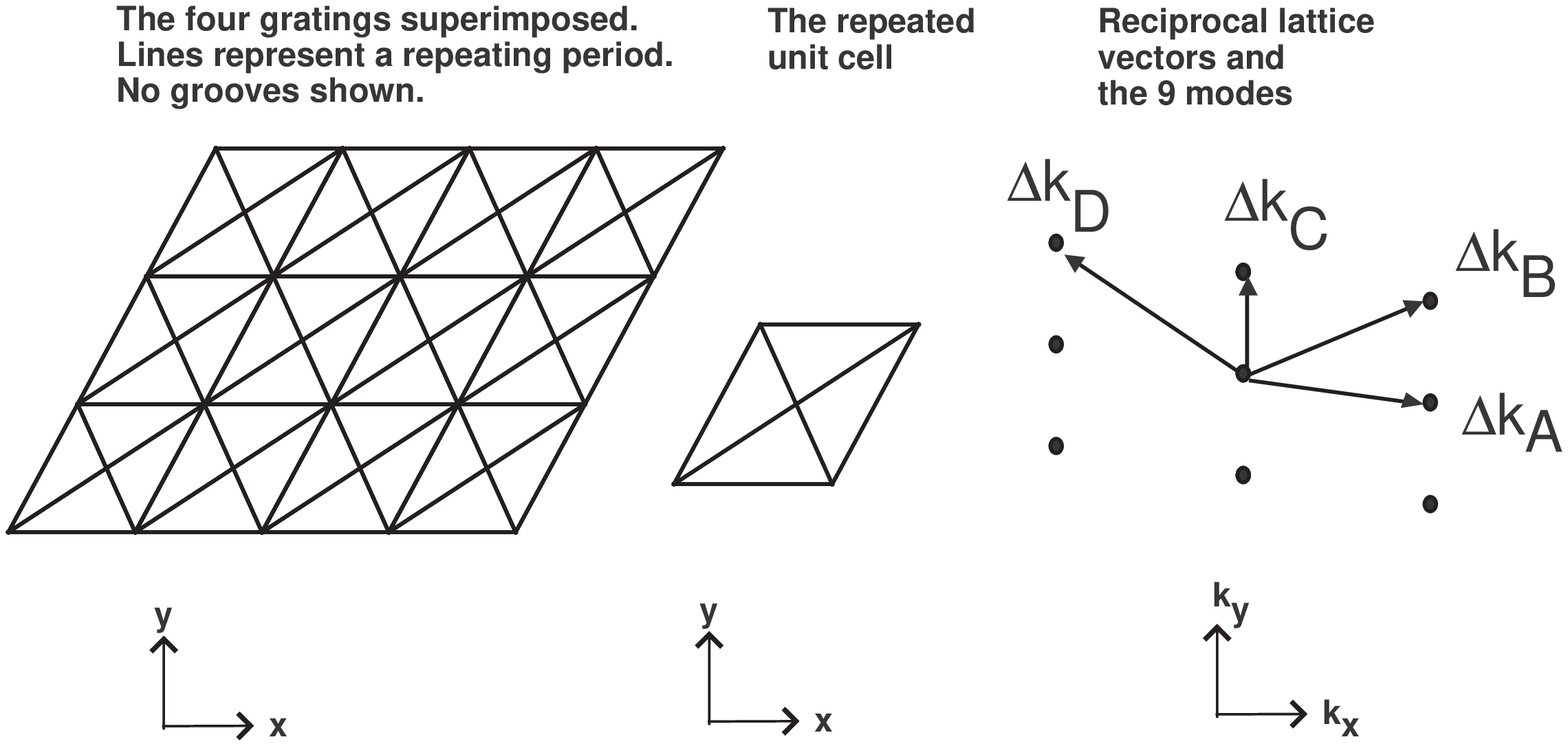}}
 \caption{\label{FigUnitCell} The choice of the grating period and
 angle for the four layers
 satisfies a repeating unit cell. This creates a finitely dense set of modes.
 The model considers only the first 9 modes.}
\end{figure}

This is the case for a one-dimensional grating.  The
two-dimensional generalization involves finding a separate $\Delta
\vec{k}_j$ for each discrete symmetry $\vec d_l$ where $j$ and $l$
enumerates the discrete symmetries.  The $\Delta \vec{k}_j$ are
called reciprical lattice vectors, and they satisfy the equation
\begin{equation}
  \Delta \vec{k}_j \cdot d_l = \delta _{j,l}
\end{equation}
where $\delta_{j,l}$ is a Kronecker delta. The reciprical lattice
vectors are given by $\Delta \vec{k}_j = \epsilon_{j,k} 2 \pi / |
\vec{d}_k| \ (\hat{z} \times \hat{g})$ where $\hat{g}$ is a unit
vector parallel to the grating grooves and $\epsilon_{j,k}$ is the
two component normalized antisymmetric tensor
$\epsilon_{11}=\epsilon_{22}=0$ and
$\epsilon_{12}=-\epsilon_{21}=1$. Because each discrete symmetry
$\vec{d}_j$ is a two-dimensional vector, we cannot find a set of
reciprical lattice vectors if there are more than two independent
discrete symmetries. Physically, more than two independent
discrete symmetries will lead to an infinitely dense set of modes
to consider.

In my structure, I need to have four or more gratings at four or
more different orientations.  However, if they are independent
discrete symmetries, I will have an infinite number of modes to
consider.  To solve this problem, I constrain the discrete
symmetries to be linearly dependent.  I limit my attention to
cases where the period and orientation of the four gratings form a
repeating unit cell as shown in fig.~\ref{FigUnitCell}. Each
grating has one reciprocal lattice vector. Because our four
gratings share a repeating unit cell, the reciprocal lattice
vectors $k_B$ and $k_D$ are formed by $\Delta k_{B (D)} = \Delta
k_C + (-) \Delta k_A$.

With this, I have a description of the modes in the bulk and a
method to propagate them using a transfer matrix.

\section{Modes in a Perfectly Conducting Grating}

\begin{figure}
 \centerline{\includegraphics[width=4.8in]{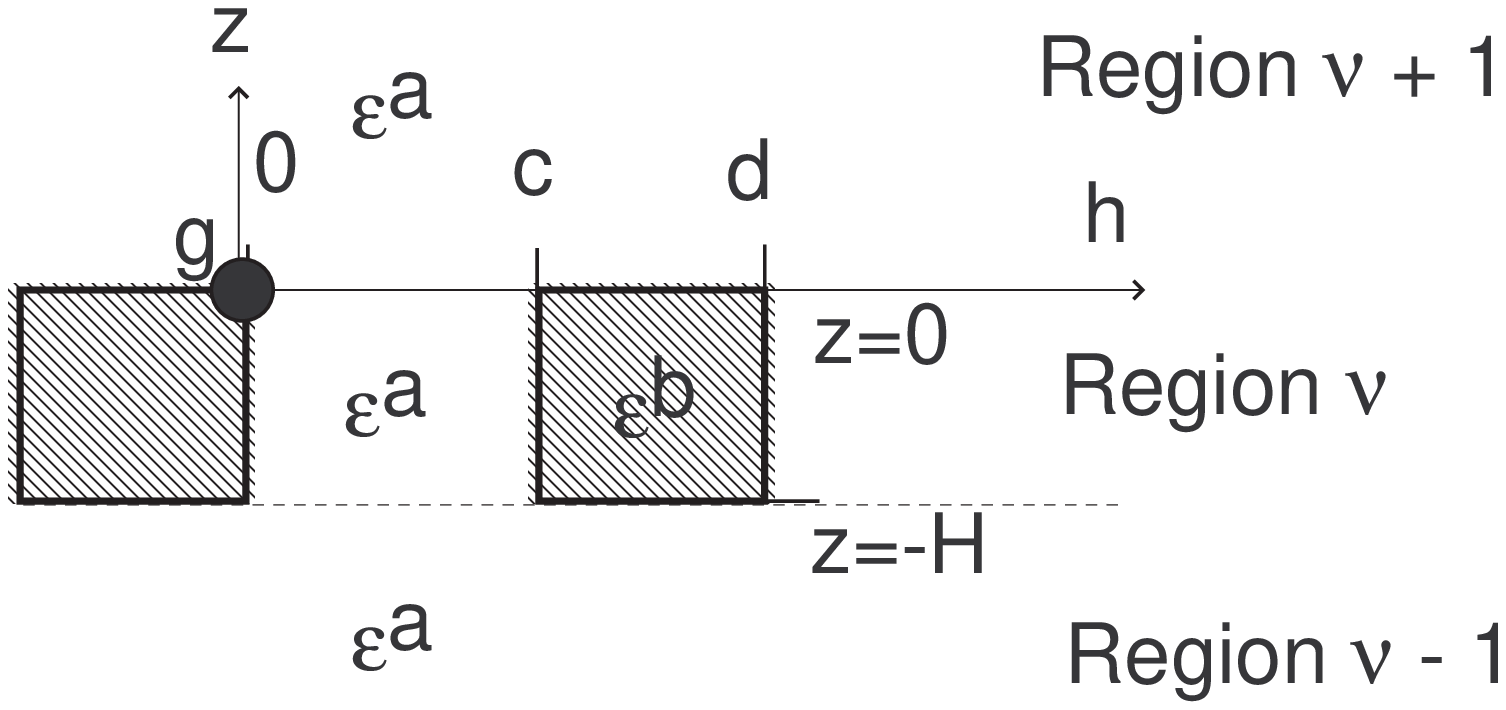}}
 \caption{\label{FigGratingCoordinateSystem} The coordinate system for the grating calculation.}
\end{figure}
The modes in a perfectly conducting, metal grating are
particularly simple. Because each grating is at a different angle,
I choose two new variables for the coordinates parallel to the
grating groove, $g$, and perpendicular to the grating grooves,
$h$. Figure \ref{FigGratingCoordinateSystem} is a diagram of the
coordinate system. The gratings are perfectly conducting for $c
\leq h \leq d$. For a grating with grooves along the g-direction,
the fast modes are
 \begin{equation}
 \Psi^{(\nu)}_{F,m,\pm} = \exp( \pm i\, \beta^{(\nu)}_m \,z + i \gamma^{(\nu)}_{m}\, g
 ) \sin (\frac{m \pi}{c}\,h)
 \end{equation}
and the slow modes are given by
 \begin{equation}
 \Psi^{(\nu)}_{F,m,\pm} = \exp( \pm i\, \beta^{(\nu)}_m \,z + i \gamma^{(\nu)}_{m}\, g
 ) \cos (\frac{m \pi}{c}\,h).
 \end{equation}

\section{Modes in a Dielectric Grating}

The modes for a dielectric grating are considerably more complex.
I found the modes are a mixture of fast and slow modes, and the
fields are represented by the waveguide modes:
 \begin{eqnarray}
 E_z & = & \frac{1}{\epsilon\, \omega^2/c^2 - \gamma^2}
 \left( i \gamma \partial_z E_g - i\frac{\omega}{c} \partial_h H_g\right) \label{WGModesFirst} \label{EqEzGeneral}\\
 E_h & = & \frac{1}{ \epsilon\, \omega^2/c^2 - \gamma^2}
 \left( i \gamma \partial_h E_g + i\frac{\omega}{c} \partial_z H_g\right) \label{EqEyGeneral}\\
 H_z & = & \frac{1}{\epsilon\, \omega^2/c^2 - \gamma^2}
 \left(  i\frac{\omega\,\epsilon}{c} \partial_h E_g +  i \gamma \partial_z H_g\right) \label{EqBzGeneral}\\
 H_h & = & \frac{1}{\epsilon\, \omega^2/c^2 - \gamma^2}
 \left( - i\frac{\omega\,\epsilon}{c} \partial_z E_g + i \gamma \partial_h H_g \right)  \label{EqByGeneral}\label{WGModesLast}
 \end{eqnarray}
To slightly simplify the expressions in this section, I have
dropped the superscript $(\nu)$ indicating which layer is being
referenced.

To solve the boundary conditions at the interfaces of constant $h$
value, the modes inside the grating must be a specific
superposition of fast and slow components.  The next several
paragraphs describe how to find the superposition of fast and slow
modes that satisfies the boundary conditions at $h=0$ and $h=c$.

The solutions in each region are expressed as
\begin{eqnarray}
\Psi_{F,m,\pm}^{(a)}=\sum_{m=1}^N  Y_{F,m+}^{a}(h)\,e^{i \gamma g \pm i \beta_m z} \ \ \ 0<h<c \ , \ -H < z < 0 \\
\Psi_{F,m,\pm}^{(b)}=\sum_{m=1}^N  Y_{F,m+}^{b}(h)\,e^{i \gamma g  \pm i \beta_m z} \ \ \ c<h<d \ , \ -H < z < 0 \\
\Psi_{S,m,\pm}^{(a)}=\sum_{m=1}^N  Y_{S,m+}^{a}(h)\,e^{i \gamma g  \pm i \beta_m z} \ \ \ 0<h<c \ , \ -H < z < 0 \\
\Psi_{S,m,\pm}^{(b)}=\sum_{m=1}^N Y_{S,m+}^{b}(h) \,e^{i \gamma g
\pm i \beta_m z} \ \ \ c<h<d \ , \ -H < z < 0
\end{eqnarray}
where
\begin{eqnarray}
Y_{X,m\pm}^{a}(h) = D^a_{X,m,R}\, e^{i \alpha_{m}^a\,h} \pm D^a_{X,m,L}\,e^{-i \alpha_{m}^a \, h} & n\,d < h< n\,d+c \  \\
Y_{X,m\pm}^{b}(h) = D^b_{X,m,R}\, e^{i \alpha_{m}^b\,(h-c)} \pm
D^b_{X,m,L}\,e^{-i \alpha^b_{m} \, (h-c)} &  n\,d+c<h < (n+1)\,d \
\end{eqnarray}
and where $X$ is $F$ or $S$ for fast or slow mode. For all the $z$
dependencies inside the grating layer, the $z$ coordinate spans
the range, $ -H < z < 0$. I also have the constraint
\begin{equation}
(\epsilon^{a}-\epsilon^{b})\,\omega^2 / c^2 = (\alpha^{a}_{m})^2
- (\alpha^{b}_{m})^2 \label{Eqalphabetaconstraint2}
\end{equation}
such that for both regions with dielectric constant $\epsilon^a$
and $\epsilon^b$, the $z$-momentum for a particular fast or slow
mode is
\begin{equation}
\beta_{m}=\sqrt{\epsilon^{(\xi)}\,\omega^2 / c^2 - \gamma^2 -
(\alpha^{(\xi)}_{m})^2}
\end{equation}
 where
$\xi=(a,b)$.

By matching boundary conditions at $y=0$ and $y=c$, I found the
following system of equations:
\tiny
\begin{equation}
M = \left( \matrix{ {\phi^a} & {\phi^{ac}} & -1 & -1 & 0 & 0 & 0 &
0 \cr 1 & 1 &
    - F\,{\phi^b}   &
    - F\,{\phi^{bc}}   & 0 &
   0 & 0 & 0 \cr - {\Gamma_h^a}\,
     {\epsilon^a}\,{\phi ^a}   &
   {\Gamma_h^a}\,{\epsilon^a}\,
   {\phi^{ac}} & {\Gamma_h^b}\,
   {\epsilon^b} & - {\Gamma_h^b}\,
     {\epsilon^b}   & - {
      \Gamma_z^a}\,{\phi^a}   & -
     {\Gamma_z^a}\,{\phi^{ac}}
   & {\Gamma_z^b} & {\Gamma_z^b} \cr -
     {\Gamma_h^a}\,{\epsilon^a}   &
   {\Gamma_h^a}\,{\epsilon^a} & {\Gamma_h^b}\,{\epsilon^b}\,F\,{\phi^b} &
   - {\Gamma_h^b}\,{\epsilon^b}\,F\,
     {\phi^{bc}}   & -{\Gamma_z^a} & -{\Gamma_z^a} & {\Gamma_z^b}\,F\,
   {\phi^b} & {\Gamma_z^b}\,F\,
   {\phi^{bc}} \cr -{\Gamma_z^a
    } & -{\Gamma_z^a} & {\Gamma_z^b}\,F\,
   {\phi^b} & {\Gamma_z^b}\,F\,
   {\phi^{bc}} & {\Gamma_h^a} & -
    {\Gamma_h^a} & - {\Gamma_h^b}\,F\,
     {\phi^b}   & {\Gamma_h^b
    }\,F\,{\phi^{bc}} \cr -
     {\Gamma_z^a}\,{\phi^a}
   & - {\Gamma_z^a}\,
     {\phi^{ac}}   & {\Gamma_z^b
   } & {\Gamma_z^b} & {\Gamma_h^a}\,
   {\phi^a} & - {\Gamma_h^a}\,
     {\phi^{ac}}   & -{\Gamma_h^b} & {\Gamma_h^b
   } \cr 0 & 0 & 0 & 0 & 1 & 1 & - F\,
     {\phi^b}   & - F\,
     {\phi^{bc}}   \cr 0 & 0 & 0 &
   0 & {\phi^a} & {\phi^{ac}} & -1 & -1 \cr  } \right)
\end{equation}
 \normalsize
where this equation satisfies
\begin{equation}
M \cdot
 \left( \matrix{
  D^a_{F,m+} \cr
  D^a_{F,m-} \cr
  D^b_{F,m+} \cr
  D^b_{F,m-} \cr
  D^a_{S,m+} \cr
  D^a_{S,m-} \cr
  D^b_{S,m+} \cr
  D^b_{S,m-} \cr
 }
 \right)=0
 \label{EqBCy}
\end{equation}
 \normalsize
 where
 \begin{eqnarray}
 \Gamma_z^a=\frac{\gamma \beta}{\epsilon^a \omega^2/c^2 - \gamma^2} \\
 \Gamma_z^b=\frac{\gamma \beta}{\epsilon^b \omega^2/c^2 - \gamma^2} \\
 \Gamma_h^a=\frac{\alpha^a \omega/c}{\epsilon^a \omega^2/c^2 - \gamma^2} \\
 \Gamma_h^b=\frac{\alpha^b \omega/c}{\epsilon^b \omega^2/c^2 - \gamma^2} \\
 F=\exp( -i\,k_{h,o} d )\\
 \phi^a = \exp( i\,\alpha^a\, c )\\
 \phi^{ac} = \exp( -i\,\alpha^a\, c )\\
  \phi^a = \exp( i\,\alpha^b\, b ) \\
 \phi^{ac} = \exp( -i\,\alpha^b\, b )
  \end{eqnarray}
and where $k_{h,o}$ is an offset momentum in the $h$ direction set
by the incident light. To find the modes, we make
$\alpha^b(\alpha^a)$ and $\beta(\alpha^a)$ both functions of
$\alpha^a$, and we scan $\alpha^a$ until we find values where
${\rm{Det}}(M)=0$. Each of these values is a mode of the grating
and enumerated with the subscript $m$. For $\epsilon^a, \epsilon^b
\in {\bf R}$, all $\alpha^a_m$ will be either purely real or
purely imaginary.

For $\gamma=0$, the matrix $M$ becomes block diagonal.  Due to
symmetry considerations, the location of the zeros of $Det(M)$ are
independent of $\gamma$.  Therefore, for compuational simplicity,
I found separately scanning the determinant of each $\gamma=0$
submatrix was faster and more reliable than finding the zeros of
the entire matrix $M$ in one scan.

For each zero we have one vector that spans the null space that
satisfy eq.~(\ref{EqBCy}). I have found that in only two cases is
the null space multi-dimensional, and in both cases the associated
mode is trivial( $E_g=0$ and $B_g=0$) and should be omitted. These
null vectors can be found using the singular value decomposition.

Finding the zeros was a major hurdle in the process.  Many curves
are like parabolas that reach down and barely cross zero.  These
situations need to return the two zeros or the resulting transfer
matrix is singular.

\section{Matching boundary conditions at slab interfaces}

Matching boundary conditions at each interface is accomplished
with the following steps:  Set up the boundary requirement in
terms of the the fields $F^{(\nu)}_{s}$,$F^{(\nu+1)}_{s}$ and the
appropriate map (eqns.~\ref{EqEzGeneral} - \ref{EqByGeneral}) from
$F$ to the vector fields of interest. Multiply both sides of the
equation by a list of complementary modes and integrating over a
unit cell. For example, the complimentary mode to $\exp( +i \gamma
g + i \alpha_m h)$ is $\exp( -i \gamma g - i \alpha_n h)$. Perform
this with enough complimentary modes to form a square non-singular
matrix. The expression will take the form of
\begin{equation}
{\mathcal{D}}^{(\nu)} C^{(\nu)}(h_\nu) = {\mathcal{D}}^{(\nu+1)}
C^{(\nu+1)}(h_\nu)
\end{equation}
where the matricies ${\mathcal{D}}^{(\nu)}$ are known as the
dynamical matricies.  This expression can be used to solve for the
transmission matrix,
\begin{equation}
D(\nu,\nu+1)= [{\mathcal{D}}^{(\nu)}]^{-1}\
{\mathcal{D}}^{(\nu+1)}.
\end{equation}

A few final notes.  The modes in the grating need not be
orthogonal because they do not satisfy the Sturm Louiville
conditions. I do not present the integrals here because I
performed them numerically in the C++ code. Currently, the
pixel-polarimeter model has about 20 design parameters and 20
parameters to describe the physics of the constituent materials
and the incident light.

\section{Calibration and Measurement Uncertainty}

Using this toolbox, the C++ code starts from a given incident
polarization state, calculates the coefficients at layer $\nu=N$,
propagates these coefficients to the regions above and below each
quantum-well stack, and calculates the energy lost in each of the
quantum-well stacks.  The absorption by the quantum-well stack is
proportional to the energy lost across the stack.   The
photocurrent is proportional to the absorption.  In this way, I
can predict the photocurrent from any number of layers given a
pure, incident polarization state. The photocurrents
representative of unpolarized light are found by averaging the
photocurrents from x-polarized light and y-polarized light.

Because the Stokes vectors are a linear vector space, the response
of an arbitrary incident polarization state can be summarized in a
polarization response matrix ($PRM$). This matrix maps the four
components of the incident light's Stokes vector $S^{\rm{in}}$ to
the photocurrents $R$:
 \begin{equation}
   R_j = \sum_{k=1}^4 PRM_{jk}\,
   S_k^{\rm{in}},
   \label{EqPRM}
 \end{equation}
where the index $j$ runs from 1 to the number of readouts $N_R$.
For full polarimetric detection $N_R \ge 4$ is required.

If $N_R > 4$,the $PRM$ represents an over-determined system. The
pseudoinverse, \emph{i.e.}~least squares fit, gives the Stokes
vector from the device readouts\footnote{I am grateful to Paul
Alsing for reminding me about the pseudoinverse in over-determined
systems.}. The Stokes vector in matrix notation with the
pseudoinverse is given by,
 \begin{equation}
   S^{\rm{in}} = (PRM^T \cdot PRM)^{-1}\cdot PRM^T \cdot R,
   \label{EqPRMInv}
 \end{equation}
where $PRM^T$ is the transpose of the matrix $PRM$. I will
abbreviate the $4 \times N_R$ pseudoinverse of the $PRM$ matrix as
 \begin{equation}
 \widetilde{PRM}^{-1} \equiv (PRM^T \cdot PRM)^{-1}\cdot PRM^T.
 \end{equation}
From the error propagation formula, the device's ability to
distinguish the distinct Stokes vectors is found  to be
 \begin{equation}
   (\delta S_j)^2 = \sum_{k=1}^{N_R} |( \widetilde{PRM}^{-1})_{j k}|^2 \ (\delta R_k)^2
 \end{equation}
where $\delta S_j$ is the uncertainty in the $j^{\rm{th}}$ Stokes
vector and $\delta R_k$ is the uncertainty in the $k^{\rm{th}}$
readout. If the uncertainty in the different readouts is equal,
the device's inherent polarization uncertainty factor is given by
\begin{equation}
  \Lambda_j=\sqrt{ \sum_{k=1}^{N_R} |( \widetilde{PRM}^{-1})_{j k}|^2 }
  \label{EqLambdaj}
\end{equation}

\begin{figure}
 \centerline{\includegraphics[width=2in]{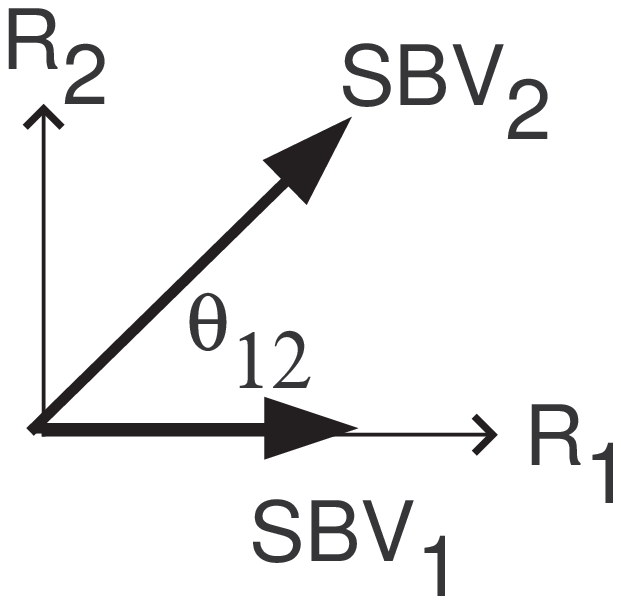}}
 \caption{\label{Fig2DBasisVectorDemo} A 2D example of two Stokes basis vectors
 ($SBV_1$ and $SBV_2$)
 in an orthonormal readout vector space ($R_1$ and $R_2$).}
\end{figure}

Conceptually, the uncertainty in $\delta S_j$ can be understood as
the ability to distinguish two Stokes basis vectors ($SBV$) on a
two-dimensional readout vector space. If the uncertainties in the
two dimensions of the readout space are assumed to be equal,
$\delta R= \delta R_1 = \delta R_2$, then the $2 \times 2$ inverse
can be calculated analytically and gives the uncertainty $\delta
S_j$ to be
 \begin{equation}
   \delta S_j = \frac{1}{\sin \theta_{12} |SBV_j|}\, \delta R
   \label{EqDeltaS}
 \end{equation}
where $\theta_{12}$ is the angle between the Stokes basis vectors
and $|SBV_j|$ is the length of the Stokes basis vector $SBV_j$ as
depicted in fig.~\ref{Fig2DBasisVectorDemo}. The uncertainty in
each Stokes vector is found to be inversely proportional to the
magnitude of that Stokes basis vector and to the sine of the angle
between that Stokes basis vector and the other three Stokes basis
vectors.

In order to relate the inherent polarization uncertainty factor
$\Lambda_j$ to a meaningful polarization detection performance, I
need to know the signal to noise ratio of our quantum-well
detectors. Using current QWIP cameras as a reference
\cite{00Pan01}, each quantum-well readout can detect changes in
photocurrent larger than about $0.02 \%$ of a room-temperature
signal (this corresponds to an $NE\Delta$$T$ of $9 mK$ for a $300
K$ blackbody background).  This gives us a signal-to-noise ratio
$SNR=R / \delta R = 5000$ for the photocurrent from each
quantum-well stack.

Using these figures, the normalized Stokes error is given by
 \begin{equation}
  \delta S_j / S_0 = \Lambda_j \ (\delta R / R) = \Lambda_j \times
  0.0002
 \end{equation}
A typical signal of interest in the infrared is 2\% to 5\%
linearly polarized.  This means that for a signal to noise ratio
of 2, we need  $\Lambda_j \leq 100$.

To simplify the data presentation, unless otherwise specified, the
subsequent plots will show the device's worst case polarization
uncertainty factor
 \begin{equation}
 \Lambda_{\rm WC}= \rm{max} \left( | \Lambda_j | \right).
 \label{EqWCPolErrorFactor}
 \end{equation}

\section{Results}

I constructed a C++ code from these elements to model the behavior
of the proposed device.  In this section, I will relate some of
the results of the model indicating the underlying physics and
ways to improve performance.

\subsection{Using Perfectly Conducting Gratings}

The predicted response of a pixel-polarimeter at $\lambda=8.42\
\mu m$ is shown in figs.~\ref{FigLinResponce} and
\ref{FigCircResponce}. In both plots, the pixel-polarimeter that
is modeled has $50$ quantum wells each with peak absorption at
$\lambda_{\rm QW}=8.42\ \mu m$, well width of $0.005 \mu m$ and
barrier width of $0.03 \mu m$.  The quantum well absorption was
modeled with a Drude model, and the Drude model parameters we used
were taken from Anderson and Lundqvist's paper
\cite{91Andersson01}. The gratings are structured with
orientations as shown in fig.~\ref{FigDevice}. The gratings at
$45^o$ and $135^o$ have periods of $3.13\ \mu m$, and, due to the
unit cell, the gratings at $0^o$ and $90^o$ have periods of
$\sqrt{2}\cdot 3.13\ \mu m = 4.42649\ \mu m$. I refer to the
period of the gratings at $45^o$ and $135^o$ as the base period.
All gratings have depths of $0.3\ \mu m$, and duty cycles with
GaAs for $0.7$ of each period followed by a perfect conductor for
$0.3$ of the period. The contact layers above and below the
gratings are $0.5\ \mu m$ each.

Fig.~\ref{FigLinResponce} shows the linear-polarization responses
of the four layers of the pixel polarimeter. The relative response
changes as the angle of the linearly-polarized incident light
varies. Fig.~\ref{FigCircResponce} shows the
elliptical-polarization response of the four layers of the pixel
polarimeter. The relative response changes as we change the phase
lag between an $x$-polarized and a $y$-polarized plane wave.
Notice that at phase lags of $0^o$ and $180^o$ the waves are
linearly polarized with angles of $45^o$ and $135^o$ respectively.

To keep these scans representative of the scattering physics and
independent of the quantum-well design choices, we set the
quantum-well peak-absorption wavelength equal to the wavelength of
the incident light.

\begin{figure}
 \centerline{
 \includegraphics[width=5in]{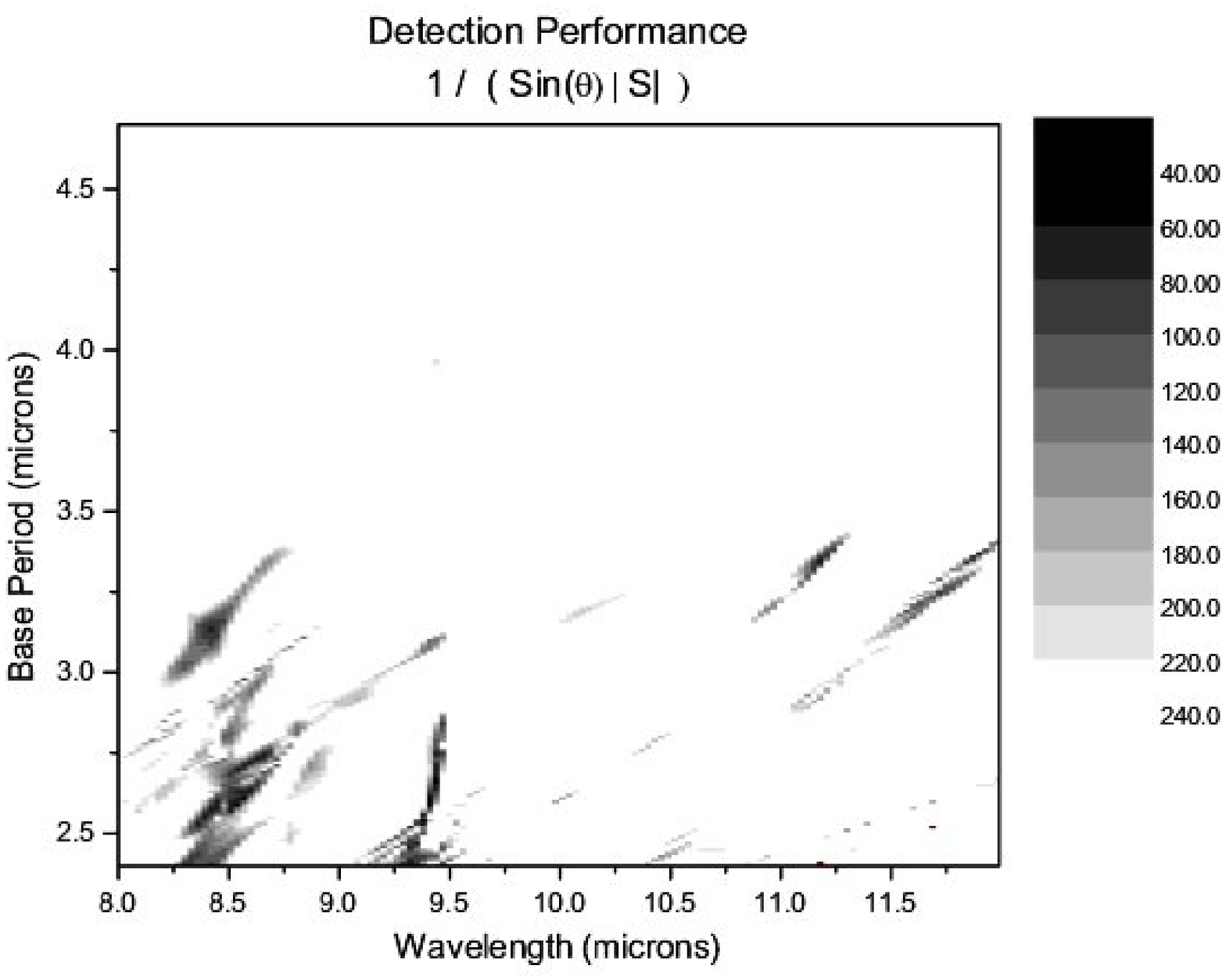}
 }
 \caption{\label{FigParamPeriod} The worst case polarization
 uncertainty factor (see equation \ref{EqWCPolErrorFactor}) for the pixel polarimeter as one varies the wavelength
 and the base period.  The lower the factor, the better the polarization detection performance.}
\end{figure}

\begin{figure}
 \centerline{
 \includegraphics[width=5in]{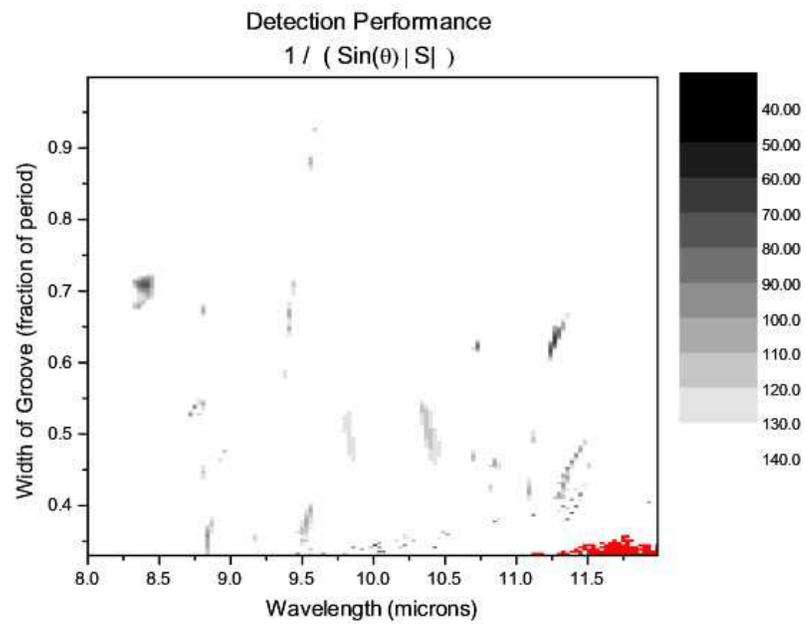}
 }
 \caption{\label{FigParamGrooveFrac} The worst case polarization
 uncertainty factor (see equation \ref{EqWCPolErrorFactor}) for the pixel polarimeter as one varies
 the wavelength and the size of the dielectric grooves
 that form the grating. The code becomes unstable for large
 wavelengths and narrow dielectric grooves.}
\end{figure}

\begin{figure}
 \centerline{
 \includegraphics[width=5in]{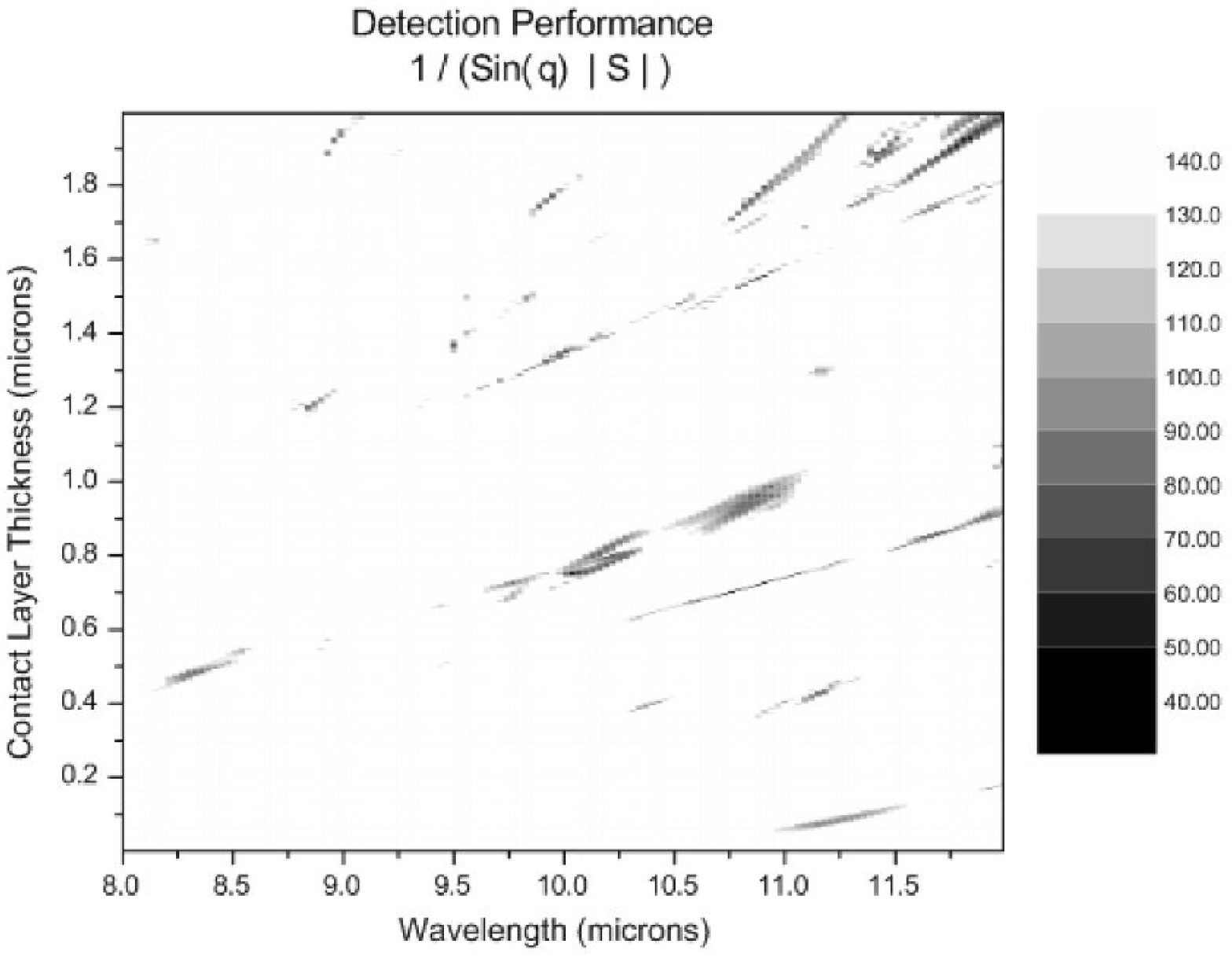}
 }
 \caption{\label{FigParamContact} The worst case polarization
 uncertainty factor (see equation \ref{EqWCPolErrorFactor}) for the pixel polarimeter as one varies
 the wavelength and the
 thickness of the contact layer between the quantum-wells and the gratings.}
\end{figure}

In figs.~\ref{FigParamPeriod}, \ref{FigParamGrooveFrac}, and
\ref{FigParamContact}, we show the regions of parameter space with
reasonable polarization detection performance.  The dark regions
correspond to parameters for a given incident wavelength where
$\Lambda_{\rm WC} \leq 100$. In these regions for some reasonable
QWIP noise and performance values, the device can detect $DOP$
changes of $0.02$.

Fig.~\ref{FigParamPeriod} shows how the polarization detection
sensitivity changes as I vary the wavelength and the base period
of the four gratings. The responses from the four layers shown in
figs.~\ref{FigLinResponce} and \ref{FigCircResponce} correspond to
the parameters of the upper-left dark region of
fig.~\ref{FigParamPeriod}.

The angled, dark streaks in figure \ref{FigParamPeriod} near 11
microns show a pattern that can be exploited in designing a
push-broom imaging spectral polarimeter. As one moves along the
focal plane of an imaging spectral polarimeter, the pixels detect
different wavelengths.  If the focal plane is designed with a
period that gets larger as the wavelength gets larger, the
resulting structure will allow near uniform polarization
sensitivity across a region of the spectrum.

Fig.~\ref{FigParamGrooveFrac} shows how the polarization detection
sensitivity changes as I vary the wavelength and the width of the
dielectric grooves in all four gratings. In
fig.~\ref{FigParamGrooveFrac} the computer simulation becomes
unstable for small groove widths and large wavelengths. Under
these circumstances, not enough light penetrates through to the
deepest MQW stack to maintain a transfer matrix that can be
meaningfully inverted.

Fig.~\ref{FigParamContact} shows how the polarization detection
sensitivity changes as I vary the wavelength and the thickness of
the contact layers above and below each MQW stack. The device's
dependence on interference between the grating layers can be
inferred from fig.~\ref{FigParamContact}.  As you vary the contact
layer thickness by $\delta$ you change the distance between the
gratings by $2\delta$ and the round-trip path-length difference by
$4\,m\,\delta$ where $m$ is a positive integer representing
interference from layer $m$. The constructive or destructive
interference is caused by two paths for light with an optical
phase difference $\Phi$ that changes with $\delta$ and $\lambda$
in the following manner:
\begin{equation}
\Phi=\frac{\sqrt{\epsilon_{\rm GaAs}}\ 2 \pi}
{\lambda}\,4\,m\,\delta=n\,\pi \label{EqIntDeterminer}
\end{equation}
where $n$ is an integer. The constructive or destructive
interference can be maintained if we change the contact layer
thickness and the wavelength in such a way that $\Phi$ is held
constant. For constructive interference features, $n$ is even. For
destructive interference features, $n$ is odd. In order to detect
the polarization, interference is needed from some round trips
between gratings at different orientations.  Full polarization
detection requires interference from all four gratings.

We can compare the slope of the trends on
fig.~\ref{FigParamContact} to Eq.~(\ref{EqIntDeterminer}) by
solving for $\delta$:
\begin{equation}
\delta= \frac{n}{8\, m \, \sqrt{\epsilon_{\rm GaAs}}} \lambda
\label{EqIntSlow}
\end{equation}
In our code, $\sqrt{\epsilon_{\rm GaAs}}=3.28$.  For interference
between the first and fourth grating, $m=3$. For contact layer
thickness around $0.5\ \mu m$ the distance between gratings is
about $2.75\ \mu m$. Recall the grating depth is $0.3\ \mu m$.
This means the actual distance traversed across the four gratings
is nearly $3.05 \mu m \times 6$. Because the wavelength of 10 $\mu
m$ radiation in GaAs is about $3\ \mu m$, the phase must advance
about  $6 \cdot \,2\pi$ during the round trip across the four
gratings. From the distance traversed, we determine the choices of
$n$ to be
 $n=12$ for
constructive interference and $n=11$ or $13$ for destructive
interference. A choice of $n=13$ and $m=3$ gives a slope of
$0.165$.  The slope of the thin line near $10.5\ \mu m$ and a
contact thickness of $0.5 \mu m$ is $0.166$.  This analysis can be
repeated for the other clearly defined lines in
fig.~\ref{FigParamContact} to discover which interference paths
dominate the polarization detection for the various wavelengths
and contact layer thicknesses.

\subsection{Using Dielectric Gratings}

The polarization sensitivity can be further improved by increasing
the fraction of light reaching the quantum wells deep inside the
pixel.  Again, to keep the scans representative of the scattering
physics and independent of the quantum-well design choices, as we
change wavelengths, we set the quantum-well peak-absorption
wavelength equal to the wavelength of the incident light.

\begin{figure}
 \centerline{
 \includegraphics[width=3.5in]{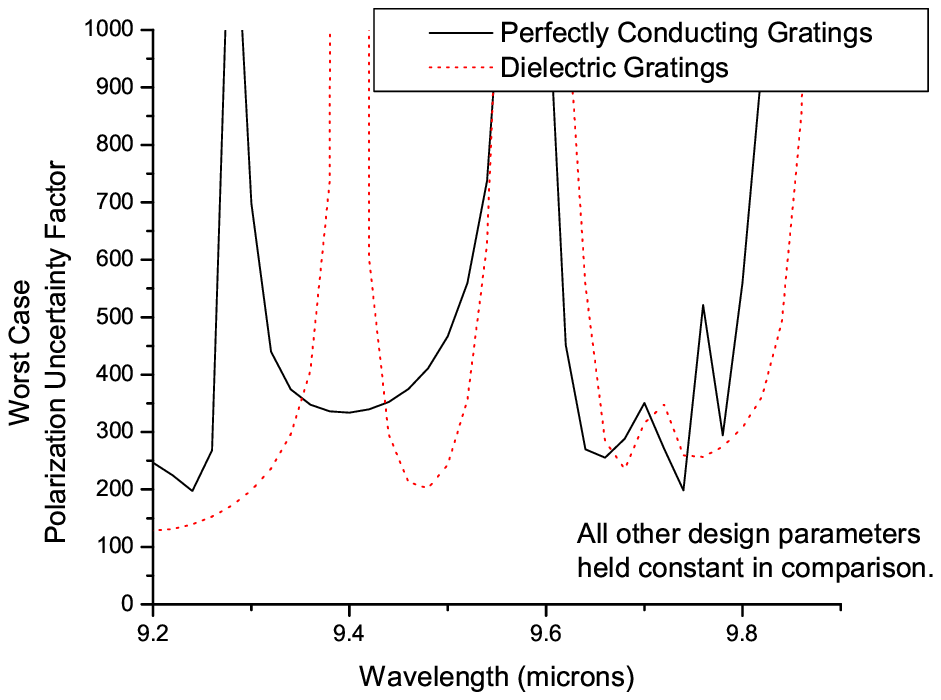}
 }
 \caption{\label{FigComparison} A representative example of
 the worst case polarization
 uncertainty factor as a function incident wavelength.  One can
 observe a general trend for the dielectric grating design to
 have lower uncertainty than the equivalent perfectly conducting grating design. }
\end{figure}

In fig.~\ref{FigComparison}, we plot the worst case polarization
uncertainty factor for a sample set of design parameters as a
function of wavelength.  Although not universal, the dielectric
grating device does have a generally lower uncertainty than the
perfectly conducting grating counterpart.  For this particular
scan, the gratings periods were the same as the scans from
figs.~\ref{FigLinResponce} and \ref{FigCircResponce}. The
dielectric grating was formed by rectangular strips of air.

The reason for the improved performance for a dielectric grating
design is that more light is able to propagate through the
$N_R$-layers of the pixel leading to a stronger signal. In the
case of perfectly conducting gratings, the stronger reflectivity
causes a very small fraction of the light to reach the deepest
quantum wells.

\section{Future modeling work}

There are many improvements that can be done with the model.
First, the code is still very unstable.  I cannot model structures
with 5 or more layers because the transfer matrices become
singular.  One way to avoid this problem is to re-write the code
with a scattering matrix approach.  Another possibility is to use
a finite time domain (FTD) model.  Although the structure is tall,
the unit cell is small and the solution should be possible in a
resonable amount of time.  This FTD approach will be too slow for
extended scans of the parameter space, but it could confirm the
presented results.

The current code has several improvements possible.  The code
currently assumes a real dielectric constant in the grating. In
practice, the GaAs is n-doped and will have some free carrier
absorption. Incorporating a complex dielectric constant for the
grating would allow us to model these effects. The modes used for
all the calculations in this section were the $3 \times 3$ set of
modes shown in figure \ref{FigUnitCell}.  I suspect the stability
would be improved if I used all modes that lie within a given
radius.  Last, the process of finding the zeros of $Det(M)$ can be
improved dramatically.  Currently, I sample the curve at a fine
interval and search for when the function changes sign.   I should
be scanning for the extrema of the curve and using these to bound
the search region when finding zeros.

With whatever code one chooses, there are also many tasks left to
be solved.  If a focal-plane array for a grating spectrometer were
to be constructed, the depth of each grating would need to remain
fixed across the focal plane.  However, something must be varied
so that the pixels will be optimized for the wavelength of the
incident light falling on that pixel.  Figure \ref{FigParamPeriod}
suggests the period could be a good choice to vary.

Also, can the device be used in conjunction with a Fourier
transform imager? Would one need to know the $PRM$ as a function
of wavelength for each pixel?  Will this perform better than the
grating spectrometer?

Last, one can always further scan or explore the 20-dimensional
parameter space to find large islands of strong polarization
sensitivity or parameters that are more easy to fabricate. I am
particularly interested in using $N_R \ge 5$ layers, rather than
the demonstrated $N_R=4$.  This will introduce an over determined
system and average out noise.  A modeling study of this approach
would be valuable.


\chapter{Process design for a two-layer polarimeter-in-a-pixel}

\label{ChapProcessDesign}

The next formidable challenge was to find a process that will
allow fabrication of the device.  We begin with a proof-of-concept
two-layer test structure.  The purpose of this structure is to
demonstrate the underlying physics of the proposed device, and to
pave the path for fabricating the more complex four-layer
structure.

The final semiconductor structure is similar to the photonic
crystal structures built by Lin, Flemming, Yamamoto, Noda,
\cite{98Lin01,98Yamamoto01,01Aoki01}. Lin and Flemming used Si and
a pattern of etchings, regrowth, and
chemical-mechanical-polishing. Aoki used InP and micromanipulation
move individual trays of periodic gratings onto a stack. Yamamoto
and Noda used GaAs and a pattern of etching gratings on separate
wafers, wafer-to-wafer bonding, and etching away the substrate.

The regrowth option is difficult when one needs to grow quantum
wells over a grating\cite{96Wernersson01}. Re-achieving
planarization is even more challenging, and the resulting quantum
wells are of low quality which degrades their responsivity and the
intersubband selection-rule critical for our application. The
micromanipulation technique is useful for fabricating gratings,
but the technique is foreign to detector development, and is not
in wide spread use even among research organizations.

I anticipate that a wafer fusion approach will be more successful
for building the pixel-polarimeter. The quantum wells will each be
grown on a perfect substrate and then gratings will be etched and
finally joined. The technique has been demonstrated on GaAs, a
favored material system for quantum-well devices and was first
developed by Dr Liau's group at MIT Lincoln Laboratory, an Air
Force FFRDC (Fedrally Funded Research and Development Center)
\cite{90Liau01}.

Because the test structure only has two layers, these pixels
cannot detect the full polarization state. The full polarization
state requires four parameters.  With a two-layer structure we can
measure at most two parameters per pixel.  This means that given
two of the Stokes parameters, we should be able to measure the
remaining two.

In this chapter, I review the techniques on which I based  my
process for the device fabrication.  I will discuss the wafer
fusion in more detail, substrate removal, and the design of the
mask set.

\section{Process overview}

The process I designed uses wafer fusion to bury quantum wells
between two gratings. We begin with a standard wafer with a thin
(300 Angstrom) AlAs etch stop layer, an n-doped contact layer, 50
repetitions of a quantum-well followed by a barrier, and topped
off with a second n-doped contact layer.  We cleave the wafer into
smaller pieces. On both pieces, we grow $Si_3N_4$ to protect the
surface. On the first piece, we etch gratings.  On the second
piece, we do not etch gratings.  We then wafer fuse the epilayers
of the first piece and the second piece together. Next, the
substrate of the second piece is removed by a selective wet
chemical etch. The sacrificial etch stop layer is also removed.
Now, gratings are etched into the stacked epilayers. We now have
two quantum-well stacks and two gratings stacked atop one another.
After the top grating, we deposit the n-metal for the top contact.

Now we isolate pixels.    We etch the top mesa, and then with a
slightly larger mask, we etch the bottom mesa. In order to make
contact to the middle mesa, we need to run a contact pad from the
bottom of the tall pixel up a wall to the middle layer of the tall
pixel.  We also need to insulate this middle contact pad from the
n-doped bottom surface. To insulate the middle contact pad from
the bottom surface, we deposit a layer of $Si_3N_4$.  The
$Si_3N_4$ coats all walls and surfaces of the structure. Holes
need to be opened in the $Si_3N_4$ to allow contact to the top of
the pixels, middle, and bottom of the pixel.  In order to ensure
electrical contact across the interface, I make separate contact
to a layer right above and right below the fused interface.

\begin{figure}
 \centerline{  \includegraphics[width=5in]{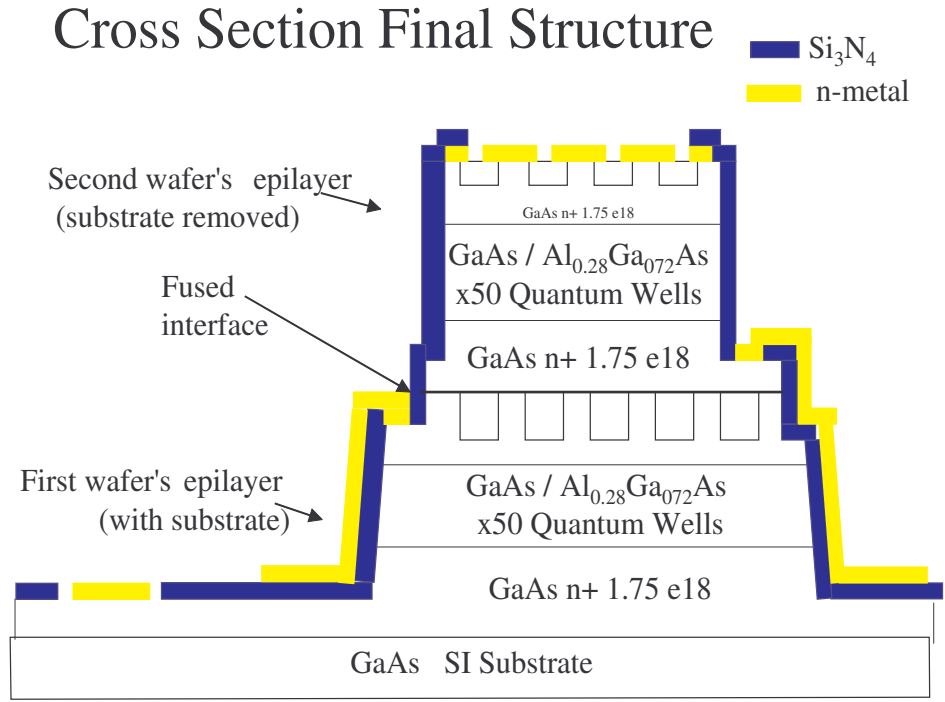}  }
 \caption{\label{FigCrossSection} A cross section of the final two-layer test structure.  }
\end{figure}
The final two-layer structure should have the cross-section shown
in figure \ref{FigCrossSection}. I have written a detailed list of
the processing steps in appendix \ref{SecProcessingStepsDetailed}.
The remainder of this section gives details and background on the
constituent steps that form this process.

\section{Wafer fusion overview}

Wafer fusion and wafer bonding are techniques used to join two
materials. Typically, but not always, wafer bonding refers to the
use of an adhesive or intermediate layer, and wafer fusion refers
to direct union of the two materials without an adhesive
\cite{98Margalit01}. Wafer bonding is an older technique and will
be skipped in this report. In addition to direct fusion and
bonding with an adhesive or intermediate layer, anodic bonding is
a third category of wafer fusion that makes use of an electric
field to encourage the bonding process \cite{98Schmidt01}.

In 1986, Lasky first reported wafer fusion to achieve a high
quality Si crystal over an insulator \cite{86Lasky01}. In 1990,
Liau and Mull from MIT Lincoln Laboratories first reported wafer
fusion in {III-V} semiconductors \cite{90Liau01}. They reported
bonding n-doped and p-doped materials and demonstrating normal
diode characteristics.  Liau and Mull also reported success on
joining materials of different lattice constants.  All their wafer
fusion was done on samples with fused areas smaller than $4\
cm^2$. Since these seminal papers, many people have achieved wafer
fusion with GaAs, InP, and Si surfaces; these examples and studies
will be addressed in the body of this chapter.

Wafer fusion is accomplished by first cleaning and polishing each
sample, placing and aligning the two wafers together in an $H_2$
environment, and then applying heat and pressure to allow the
bonds to break and reform across the interface.

The wafer surfaces to be fused need to be smooth and clean to
enable wafer fusion to take place.  The surfaces can to have a
local RMS roughness of no greater than about 10 Angstroms, and the
overall wafer surface must have bowing of less than 5 $\mu m$ per
4 $in$ wafer \cite{98Schmidt01}.

In some cases, one of the layers can be prepared with shallow
channels to allow liquid and gas bubbles to escape and improve the
uniformity of the fused surface \cite{98Margalit01}. Another
alternative is to include gratings or other patterns that allow
bubbles to percolate to these collection points \cite{90Liau01}.

After the wafers are cleaned and joined, they are heated under
pressure.  The device that performs this process is called a wafer
fusion reactor.

The original paper by Liau \emph{et.al.} on wafer fusion, used a
wafer-fusion reactor based on the differing coefficients of
thermal expansion between graphite and quartz \cite{90Liau01}.  A
similar setup was used by ref.~\cite{97Zhu01}.  Others have found
the pressure required is not that large; conventional pressure
source can also work. The setup used at University of California,
Santa Barbara (UCSB) is based on a half dome pressing the two
wafers together.  The pressure source is tightening screws that
hold the dome in place. The dome is critical because it diminishes
the dependence on tightening the screws evenly
\cite{98Margalit01}.

Wafer fusion is facilitated by heat and uniform pressure. The
pressure brings the two surfaces into contact and eliminates air
bubbles \cite{98Schmidt01}.  The heat allows the crystal bonds to
re-arrange and join the two surfaces.  The wafer fusion reactor is
designed such that the two wafers experience nearly uniform
pressure. The pressure used in wafer fusion varies from 3 kPa to 3
MPa \cite{97Black01}. Higher pressure should allow fusion at lower
temperatures.

\section{Substrate Removal Overview}

One of the dominant reasons to use wafer fusion is to transfer the
epitaxial layer of one wafer onto the epitaxial layer of the
second wafer. Once wafer fusion has been performed, the substrate
of one of the wafers needs to be removed
\cite{97Zhu01,98Margalit01}. This is facilitated by the use of a
pregrown, thin, etch-stop layer on the substrate to be removed.

One problem not clarified in the literature is how to protect the
back side. One method that we considered was to spin photoresist
on the back.  Another possibility is to mount the samples on a Si
wafer or a glass slide with wax.  As will be discussed in the next
chapter, the glass slide technique was found to be better.

To remove GaAs substrates, Zhu suggests jet etching with
$NH_4OH$-$H_2O_2$ solution with a PH of 8.4. During jet etching,
protect the wafer edges with wax to prevent undercut. The etching
will stop at any layer with $Al_xGa_{1-x}As$ ($x>0.6$) or a
lattice matched InGaP layer.  Both InGaP and AlGaAs etch stop
layers can be removed with HCl.

The GaAs vs AlAs selective etching technique which was best
documented was a citric acid : $H_2O_2$ etch
\cite{91Tong01,98Kim01,97CarterComan01}. Citric-acid monohydride
is mixed 1:1 with deionized water to form a citric-acid solution.
The citric-acid solution is then mixed with $H_2O_2$ at a ratio of
4:1 to form the etching solution. The solution has a selectivity
for etching GaAs over AlAs of over 1000 to 1.

As explained in the next chapter, we deduced a few missing details
not mentioned in the literature.  The etch rate was not
sufficiently uniform to remove the entire 500 micron substrate
with just the wet etch, so we needed to lap the substrate down to
100 microns.  We also found the solution looses potency after a
couple of hours so the solution needed to be replaced several
times during an etching procedure.  The end-point is easy to tell
because the rough surface of the etched GaAs becomes a smooth
glass like surface when the etching reaches the AlAs.  Last the
etch-stop layer should not be too thick because the AlAs forms a
stressed oxide when exposed to the solution
\cite{97CarterComan01}.  The stressed oxide will crumble thin
samples.

\section{A map of the mask set}

\begin{figure}
 \centerline{  \includegraphics[width=3.5in]{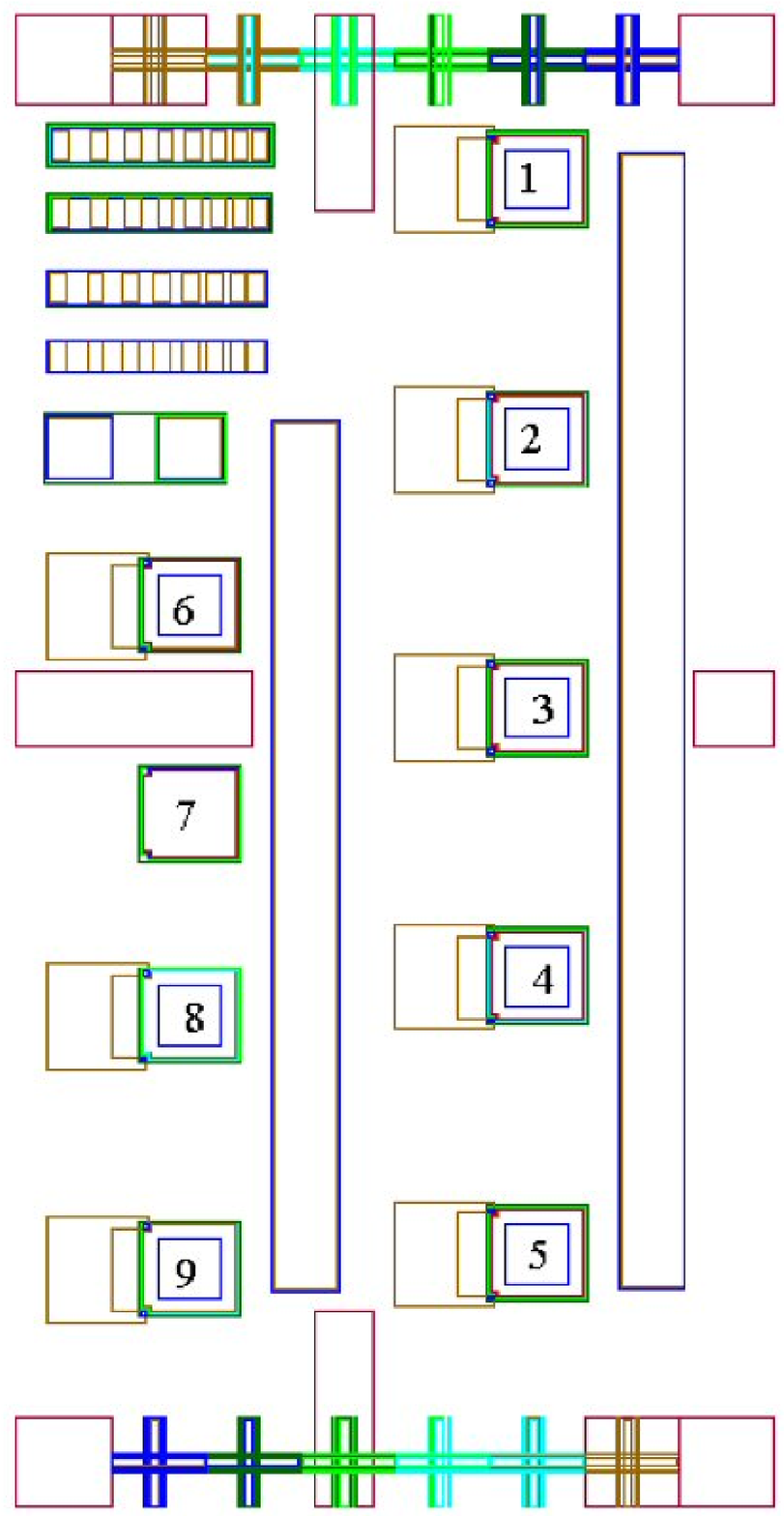}  }
 \caption{\label{FigMaskOverview} A CAD schematic of a die in the mask set that I designed.  }
\end{figure}
With these processes in mind, I designed the mask set using AutoCAD software.

Each die contains six pixels optimized to distinguish between two Stokes parameters
given the other two Stokes parameters are known.  Each die also contains a number of
diagnostic structures.  The die is shown in figure \ref{FigMaskOverview}.

\begin{figure}
 \centerline{  \includegraphics[width=3.5in]{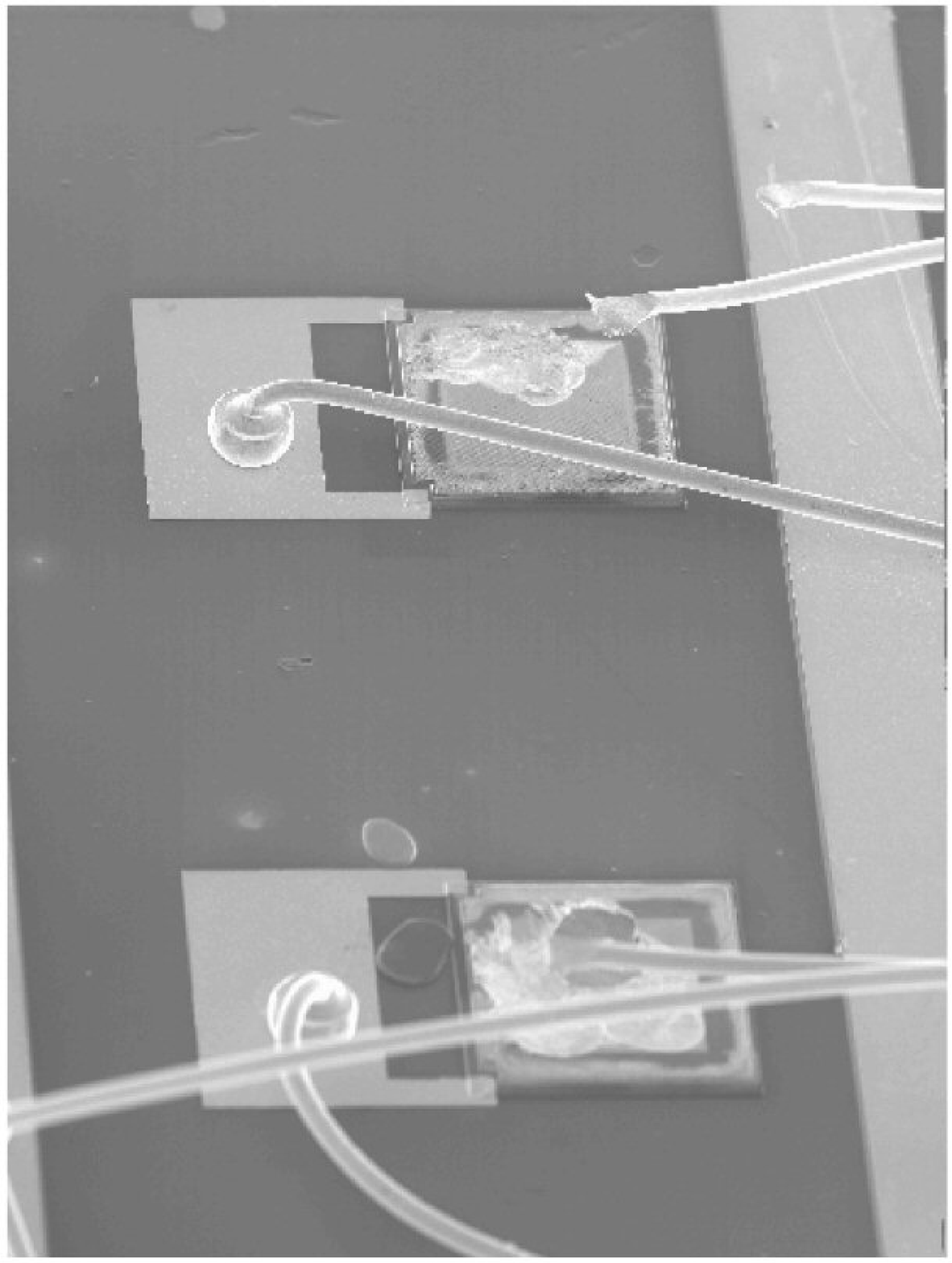}  }
 \caption{\label{FigSEMContactPads} An SEM of the three contact pads and the two-layered pixel.  }
\end{figure}
Each two-layer pixel on the die allows us to make contact to the
top, middle, and bottom of the pixel with separate wire bonds.
Figure \ref{FigSEMContactPads} shows an SEM of the contact pads to
the three layers of the pixel from our first round of fabrication.
On top of each pixel, we have a top contact pad. The `C' shaped
structure to the left of the pixels is the middle layer's contact
pad. The middle contact pad has metal arms that climb up to the
mesa and make contact with the middle of the structure. The long
strips to the right of the column of pixels form a bottom contact
pad. Each contact pad is about 200 to 250 microns.  The large size
was to enable easy wire bonding.

The right column contains pixels 1 through 5 which are each
designed to be sensitive to distinguish two polarization states.
Pixel 8 on the left column is also designed to distinguish
polarization states.  Table \ref{TablePixelParameters} lists the
design parameters of these six pixels and their intended
polarization sensitivity.  The next section explains how I chose
these design parameters and how sensitive the performance is to
variations in construction.

In practice, the fabrication process will probably not be able to
guarantee the design parameters to allow predictive design of the
structure. I expect that the irregularities in the manufacture
process will primarily shift the optimal wavelength for the
polarization sensitivity.

The left column contains mostly diagnostic devices. The top four
horizontal bars are mesas with small contact pads at varying
distances from each other.  I will refer to these as ohmic contact
probing mesas. There is one such horizontal mesa for each depth
that ohmic contacts are to form.  These structures will allow us
to measure the quality of the ohmic contacts.

Below the ohmic contact probing mesas, two small squares side by
side will test the electrical behavior of the fused interface.
The left square is just below the fused interface; the right
square will be just above the fused interface.  Measuring the
resistance across these two pads will give some indication of the
resistance across the fused interface.

Pixels 6 and 7 are to test the quantum wells. If we have
difficulty with the main experiment (pixels 1-5 and 8), these two
structures will allow us to test if the quantum wells in either
layer are working, and if so which layer is providing the
photocurrent.  This could also be a useful way of measuring the
amount of free-carrier absorption in the doped region between the
two quantum-well stacks.  Pixel 6 has no gratings at the fused
interface, and a grating at the very top of the structure.
Photocurrent can be generated and measured across both sets of
quantum-well stacks. Pixel 7 has had the top quantum-well stack
etched away and consists only of the lower quantum-well stack and
a grating on the same layer as the fused interface.

Pixel 9 is a copy of pixel 4 with one change.  The gratings extend
to the edge of the mesa.  There was no clear answer or reason that
the gratings should be exposed to the air or trapped in the
semiconductor structure.   This device explored this variation.  I
also hoped this device would expose the fused gratings to
facilitate good SEM pictures.

Last, there are several large rectangles on the edge of the die.
These are windows to facilitate aligning the metal-field masks.
They were also intended as visible alignment marks to be used in
the first alignment step after the substrate removal.

\section{Choosing the grating design parameters}

\begin{figure}
 \centerline{  \includegraphics[width=4.5in]{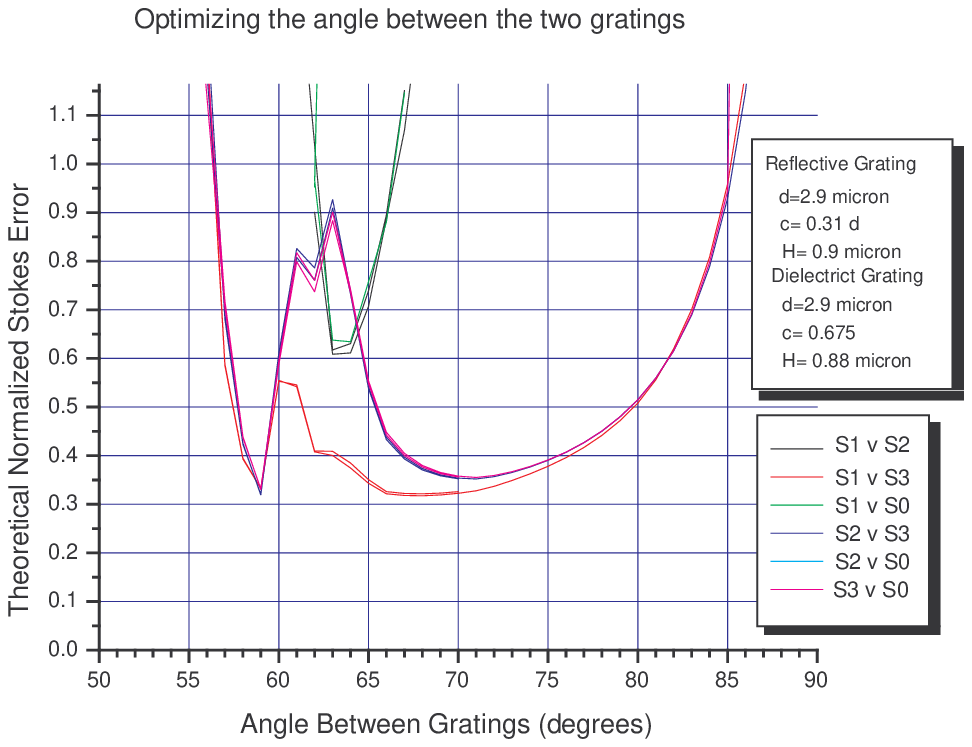}  }
 \caption{\label{FigAngleOpt} Worst case $\delta S / S_0$ for fixed grating parameters as we change the
 angle between the two gratings. }
\end{figure}
\begin{figure}
 \centerline{  \includegraphics[width=5.0in]{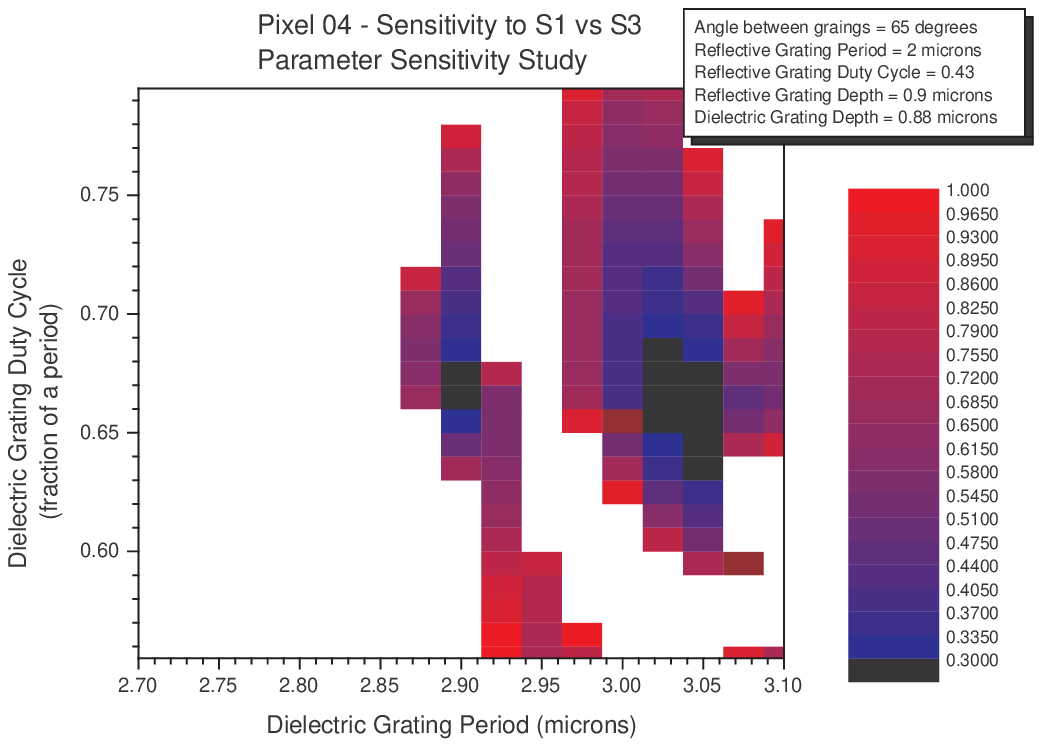}  }
 \centerline{  \includegraphics[width=5.0in]{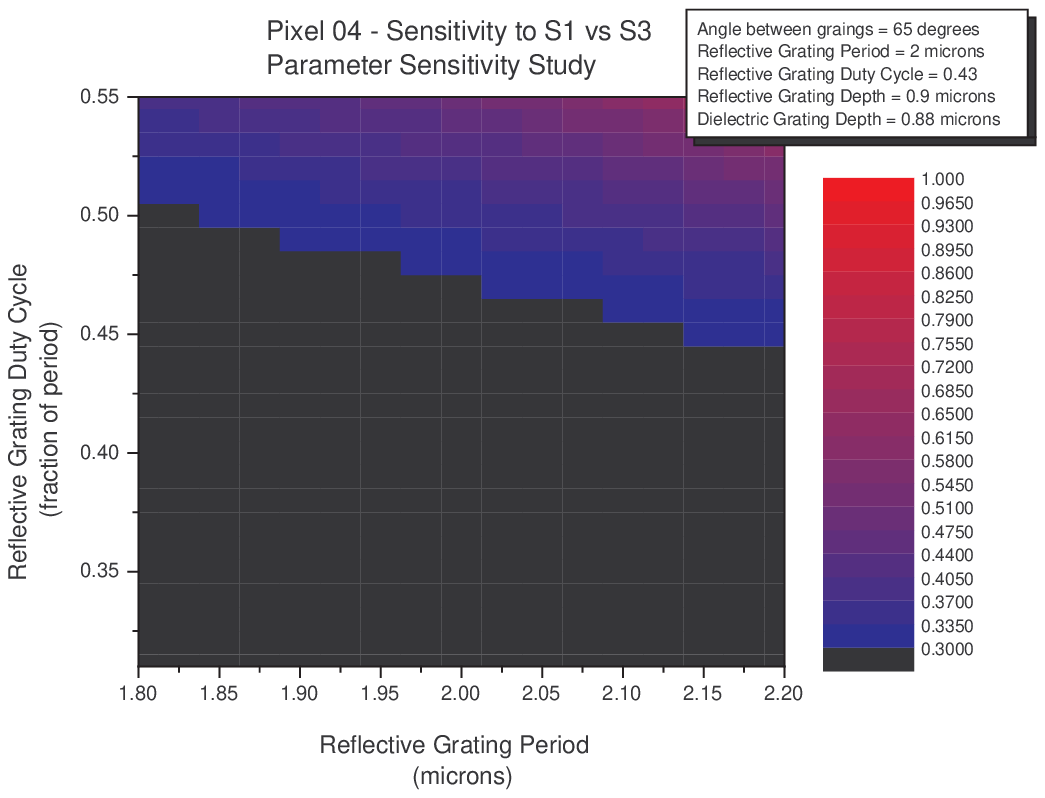}  }
 \caption{\label{FigParamSensitivity} Worst case $\delta S / S_0$ for fixed wavelength as
 I change the angle between the two gratings. }
\end{figure}

The computer model described earlier was modified to calculate the
polarization sensitivity of a two-layer structure.  I scanned
parameter space to find parameters that give reasonable
performance and designed pixels on the mask to these
specifications. During this process, I encountered substantial
instabilities in my computer code.

As I changed the angle between the gratings, energy conservation
was off by more than 100\%.  After weeks of debugging, I found the
instability was due to my choice of modes hard wired into the
code. The code was relatively stable for gratings whose angular
separation is greater than 50 degrees.  If I were able to choose
modes that lie within a fixed radius, I expect this instability
would be substantially more contained.

To account for this numerical instability, I added the numerical
error to the normalized Stokes error to create a total error
measure. The numerical error was estimated by the loss of energy
conservation. To measure energy conservation with my code (as
written), I needed to assume a lossless contact layer (no
free-carrier absorption). The numerical error was added in
quadrature to the inherent normalized Stokes error.  All plots
shown in this section include both sources of error.

The decision to model without free-carrier absorption was found to
be reasonable in several ways.   For parameters found to be
stable, I noted that the free carrier absorption lessened the
signal by a small amount and shifted the wavelength of peak
polarimetric sensitivity.  This suggested that the design
parameters found with the free-carrier absorption turned off would
be good for detecting polarization at some wavelength near the
design wavelength, but not at the design wavelength.  When I say
near, I mean within 0.1 microns of the 9 micron design wavelength.

Figure \ref{FigAngleOpt} shows a scan across the angular
separation between the two gratings while keeping all other
grating parameters constant.  Only normalized Stokes error values
less than one can be trusted to give a lab-observable polarization
sensitivity . The different curves represent the ability to
distinguish between two polarization states, for example $S_1$ and
$S_2$.

How closely does the fabrication process need to reproduce these
design parameters?  Figure \ref{FigParamSensitivity} provides us
with some indication.  In this figure, black means good
polarization sensitivity and white is unacceptable polarization
sensitivity. For the dielectric gratings in the top plot, the
period must be specified to within 0.03 microns.  The duty cycle
(grating groove width / grating period) has more flexibility. The
reflective grating parameters show much greater tolerance. The
entire scanned region of parameter space is acceptable. The
performance appears to improve as we move to shorter periods and
smaller duty cycles. This rapidly becomes impossible to fabricate
due to very small feature sizes.  I limited the feature sizes on
the mask set to 0.85 microns.

To choose these design parameters, I ran hundreds of scans on 10
computers over the course of two weeks generating nearly 2
gigabytes of data.  The results are summarized in table
\ref{TablePixelParameters}.

During these scans, two classes of solutions become apparent. The
first class involved both gratings diffracting incident light at
angles in the range of 50 degrees to 80 degrees.

The second class of solutions had one grating with a period such
that the light was diffracted to some angle in the range of 50
degrees to 80 degrees.  The second grating had a period so small
that there were no diffracted orders. The grating's purpose was to
create a bi-refringent layer to rotate the polarization state of
light diffracted by the first  grating.

Although both types of solutions displayed sensitivity similar to
that summarized in figure \ref{FigParamSensitivity}, the solutions
where one grating was behaving purely like a birefringent layer
were much more tolerant to changes in the design parameters.

\begin{table}
 \centering
  \begin{tabular}{|p{0.5in}|p{1.8in}|p{1.8in}|p{1.3in}|}
   \hline
   $\ $ & $\ $ & $\ $ & \\
    \parbox{0.5in}{\textbf{Pixel Num}}
    & \parbox{1.7in}{\textbf{Reflective Grating Parameters}}
    & \parbox{1.7in}{\textbf{Dielectric Grating Parameters}}
    & \parbox{1.2in}{\textbf{Predicted Performance Information}} \\
       $\ $ & $\ $ & $\ $ & \\ \hline
   $\ $ & $\ $ & $\ $ & \\
    1
    & \parbox{1.7in}{
        Angle : 148.9 degrees \\
        Period: 2.81 microns\\
        Groove/Period: 0.302 \\
        }
        &
        \parbox{1.7in}{
        Angle : 32.3 degrees \\
        Period: 2.91 microns\\
        Groove/Period: 0.66 }
        &
        \parbox{1.2in}{
        For $S_1$ vs ($S_2$,$S_4$) \\
        $\delta S / S_0 \leq 0.37 $.
        } \\
            & & & \\ \hline
   $\ $ & $\ $ & $\ $ & \\
    2
    & \parbox{1.7in}{
        Angle : 145.6 degrees \\
        Period: 2.92 microns\\
        Groove/Period: 0.33 \\
        }
        &
        \parbox{1.7in}{
        Angle : 35.7 degrees \\
        Period: 3.02 microns\\
        Groove/Period: 0.66 }
        &
        \parbox{1.2in}{
        For $S_3$ vs $S_x$ \\
        $\delta S / S_0 \leq 0.27 $.
        } \\
            & & & \\ \hline
   $\ $ & $\ $ & $\ $ & \\
    3
    & \parbox{1.7in}{
        Angle : 160.9 degrees \\
        Period: 2.0 microns\\
        Groove/Period: 0.43 \\
        }
        &
        \parbox{1.7in}{
        Angle : 28.4 degrees \\
        Period: 2.9 microns\\
        Groove/Period: 0.67 }
        &
        \parbox{1.2in}{
        For $S_1$ vs Sx \\
        $\delta S / S_0 \leq 0.27 $.
        For $S_3$ vs ($S_2$,$S_0$) \\
        $\delta S / S_0 \leq 0.45 $.
        } \\
            & & & \\ \hline
   $\ $ & $\ $ & $\ $ & \\
    4
    & \parbox{1.7in}{
        Angle : 154.2 degrees \\
        Period: 2.0 microns\\
        Groove/Period: 0.43 \\
        }
        &
        \parbox{1.7in}{
        Angle : 39.1 degrees \\
        Period: 3.05 microns\\
        Groove/Period: 0.66 }
        &
        \parbox{1.2in}{
        For $S_3$ vs $S_x$ \\
        $\delta S / S_0 \leq 0.22 $.
        } \\
            & & & \\ \hline
   $\ $ & $\ $ & $\ $ & \\
    5
    & \parbox{1.7in}{
        Angle : 149.6 degrees \\
        Period: 2.8 microns\\
        Groove/Period: 0.3 \\
        }
        &
        \parbox{1.7in}{
        Angle : 33.1 degrees \\
        Period: 3.04 microns\\
        Groove/Period: 0.67 }
        &
        \parbox{1.2in}{
        For $S_2$ vs $S_1$ \\
        $\delta S / S_0 \leq 0.45 $.
        For $S_2$ vs $S_0$ \\
        $\delta S / S_0 \leq 0.23 $.
        } \\
            & & & \\ \hline
   $\ $ & $\ $ & $\ $ & \\
    8
    & \parbox{1.7in}{
        Angle : 90 degrees \\
        Period: 2.92 microns\\
        Groove/Period: 0.31 \\
        }
        &
        \parbox{1.7in}{
        Angle : 0.0 degrees \\
        Period: 3.05 microns\\
        Groove/Period: 0.67 }
        &
        \parbox{1.2in}{
        For $S_1$ vs $S_0$ \\
        $\delta S / S_0 \leq 0.2 $.
        } \\
            & & & \\ \hline
  \end{tabular}
  \caption{\label{TablePixelParameters} Pixel Design Parameters.  All the reflective gratings have a depth of 0.9 microns,
  and all the burried dielectric gratings have a depth of 0.88 microns.
  The groove / period ratio uses the groove seen by the light.  This groove is the compliment to the groove one etches during processing.}
\end{table}


\chapter{Lessons from the first round of fabrication}
\label{ChapLessons}

The story of fabricating our two-layer test device reads like a
Greek epic / trajedy. Our heroes leave home for a long journey
away from family to prove their worth, gain fame and fortune, and
bravely overcome nearly-insurmountable odds. They narrowly escape
near-certain destruction on many occasions. After months away on
the long voyage, they return home and tell the glorious stories of
their battles to their friends and family. Years later, their
tales inspire the next generation of adventures to leave the nest
and brave uncharted waters.

In this chapter, I will describe the journey that we took to build
a first iteration of a two-layer proof-of-concept device. I
describe the several Achilles' heels of the proposed process, and
I will give my suggestions to avoid these obstacles.  I will also
describe the near-certain destruction that we found on other
paths. After all, the drama and intrigue of the story lie in the
battles and challenges that we encountered.  At the end of this
story, we find the working device still not in our possession but
within sight. The reader is left with one burning question: Will
there be a sequel?

\section{The plan}

During the time between November 2002 and February 2003, I was
trying to isolate some numerical instabilities in the computer
model, and I was planning how to fabricate the structure.  The
largest obstacle at the time was the wafer-fusion.  The University
of New Mexico has not performed wafer fusion with GaAs and did not
have the facilities that the literature described for the process.

During October of 2002, I tried to collaborate with Prof. Noda
from the University of Kyoto, but the international issues proved
too complex.  During November and December of 2002, I tried
collaborating with Sandia.  Although this looked promising, my
contact became too busy.  In February of 2003, I established a
collaboration with MIT Lincoln Laboratory to do the wafer fusion.

The University of New Mexico would ship Dr.~Liau at Lincoln Labs
the samples to be fused, and they would ship them back fused. Dr
Liau did request that $Si_3N_4$ be deposited before performing any
photolithography or etching to protect the surface.  UNM would
perform the remainder of the processing including substrate
removal, etching the mesas, $Si_3N_4$ deposition, the metal
contacts, and finally the characterization.

The masks were designed by early March 2003.  Processing was set
to begin the last week of March 2003.  Purchasing paperwork
delayed mask delivery until early April.  The original sample
grown for this device was found unsatisfactory so a new sample was
grown in mid April.  An FTIR test was performed that week to
verify the new QW structure.  Sample processing began the last
week of April.

The processing challenges described in this chapter continued to
slow us down.  The wafers were shipped to MIT Lincoln laboratory
on 6 May 2003.

The wafers were fused by 8 May 2003, but paper work to have them
shipped delayed their arrival until 15 May 2003.  During that
week, I debugged the substrate removal process.  By 19 May the
substrate of the fused sample was removed, and we began what we
thought would be standard processing steps.

On 22 May, we discovered the difficulty with the wet etch for the
mesa.  Addressing this issue delayed us until the beginning of
June.  During June, we debugged the $Si_3N_4$ etch and had a
finalized sample wire bonded and ready to characterize on 10 June
2003.

Developing a characterization setup was done in parallel to the
sample processing during the months of May and June 2003.  The
polarimetric tools needed for characterization were borrowed from
Tom Caudill's research group at AFRL/VS.  The characterization
setup was tested on well-characterized QWIP structures with linear
gratings.

With the story in place, I now delve into details of each
encounter with Murphy's law.

\section{The quantum-well structure}

Two wafers were grown for this device.  The first wafer was UNM
1374.  Two QWIPs were processed from this device.  Each QWIP
exhibited a short like behavior at 77 Kelvin.   The dark current
was in the range of a mAmp at 1 to 2 volts of bias.  There was no
discernible photocurrent above this high dark current.

After comparing the structure of 1374 to structures designed at
the Air Force Research Laboratory, Space Vehicles Directorate
(AFRL/VS) by Dr. Dan Huang, we noticed the quantum-well doping was
significantly higher in the 1374 sample. We hypothesize that the
Fermi-level was raised so high that the upper quantum well state
was largely occupied even at 77K.  This would explain the 'short'
like behavior of the two devices processed from this wafer.

\begin{figure}
 \centerline{  \includegraphics[width=5.0in]{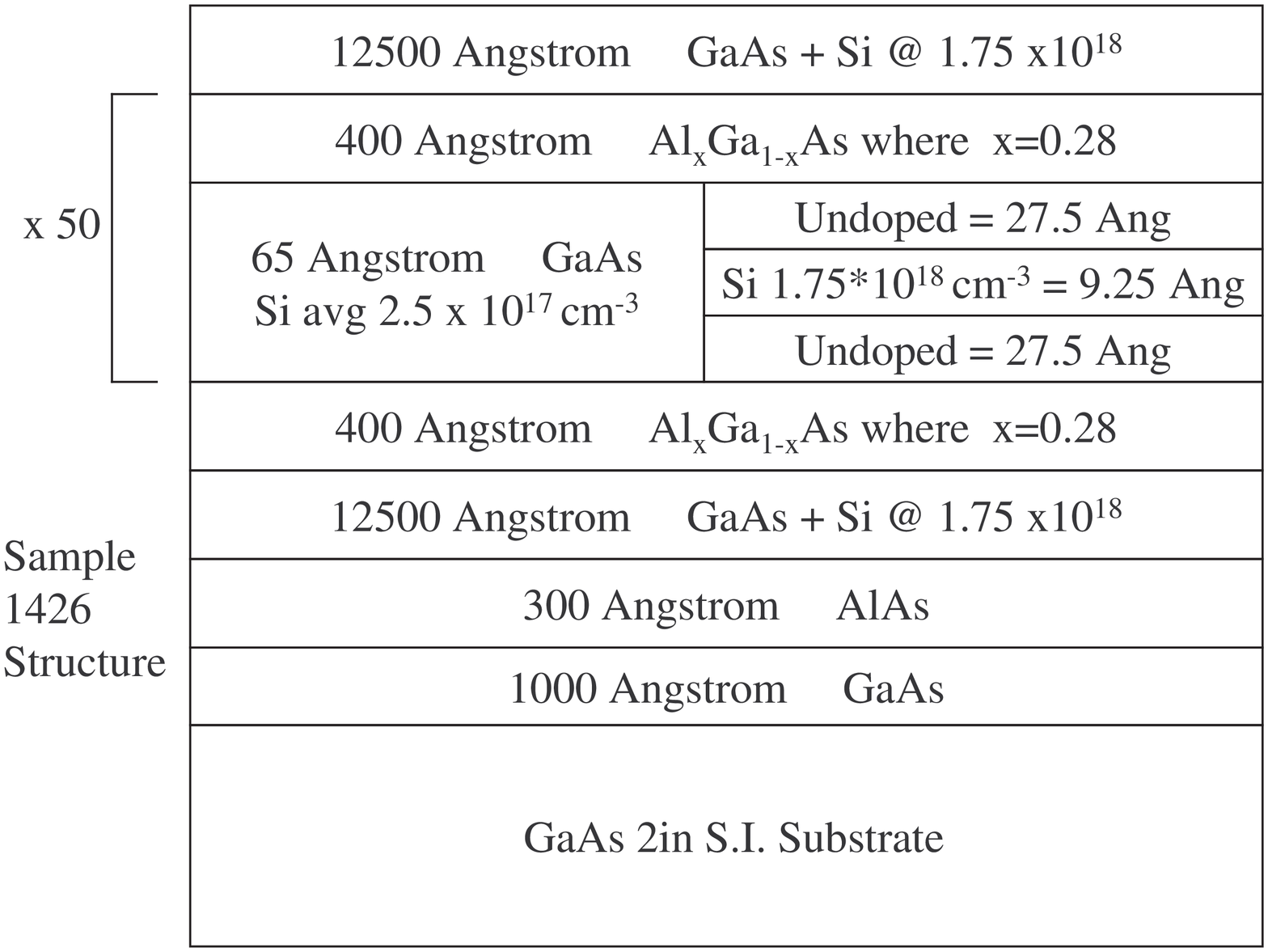}  }
 \caption{\label{FigWaferEpitaxy} The structure of UNM wafer 1426. }
\end{figure}

We designed our next structure more closely to the validated
designs used at AFRL/VS.  The main changes were to design the
sample for 9 micron absorption and to use a contact-layer doping
of $1.75 \times 10^{18} cm^{-3}$ instead of the $2 \times 10^{18}
cm^{-3}$ doping. The 9 micron absorption was because this was the
wavelength for which the mask set was already designed.  The
reasoning to lower the doping was to mitigate excess free-carrier
absorption without sacrificing the ability to form ohmic contacts.
The new structure is shown in figure \ref{FigWaferEpitaxy}.

\begin{figure}
 \centerline{  \includegraphics[width=5.0in]{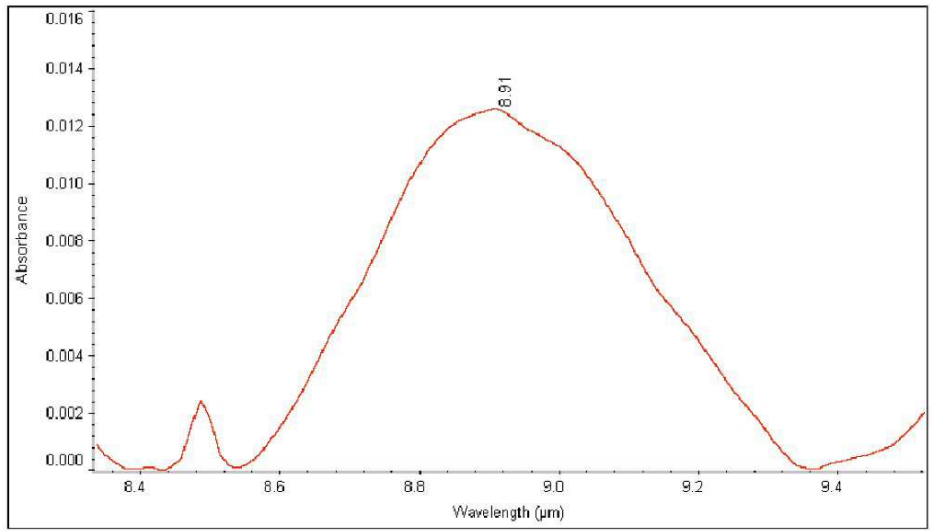}  }
 \caption{\label{FigWaferAbsorbance} The absorbance spectra of sample 1426 measured with an FTIR at Brewster's angle. }
\end{figure}

To verify the quality of the new sample in a more timely manner
than processing a single-layer QWIP, we measured the sample's
absorption spectra with an FTIR at Brewster's angle.  As can be
seen in figure \ref{FigWaferAbsorbance}, the sample showed an
absorption near 9 microns.

As will be described later in this chapter, our device did not
form good ohmic contacts.  To test if this low doping level is the
cause, I suggest processing a more simple structure and verifying
that ohmic contacts can be formed.

\section{Etching the gratings}

Patterning the photoresist for the gratings was much more
difficult than anticipated.  Per the suggestion of Beth Fuchs, we
used AZ 5206 photoresist.  Because we have features just under one
micron, the thin photoresist should be able to transfer the
pattern with greater ease.  There were three main complications,
an unknown process, poor adhesion to the $Si_3N_4$, and a
different etch rate in the gratings due to transport limitations
in the ICP.

\begin{figure}
 \centerline{  \includegraphics[width=4.0in]{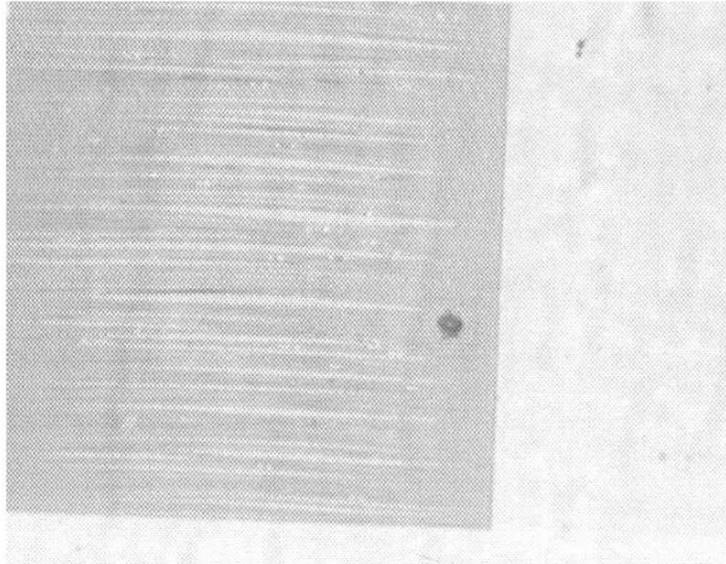}  }
 \caption{\label{FigSpagettiPR} For a variety of conditions, despite using HMDS, the photoresist AZ5206 would loose adhesion
 to the GaAs or $Si_3N_4$ surface. }
\end{figure}

\begin{figure}
 \centerline{  \includegraphics[width=4.0in]{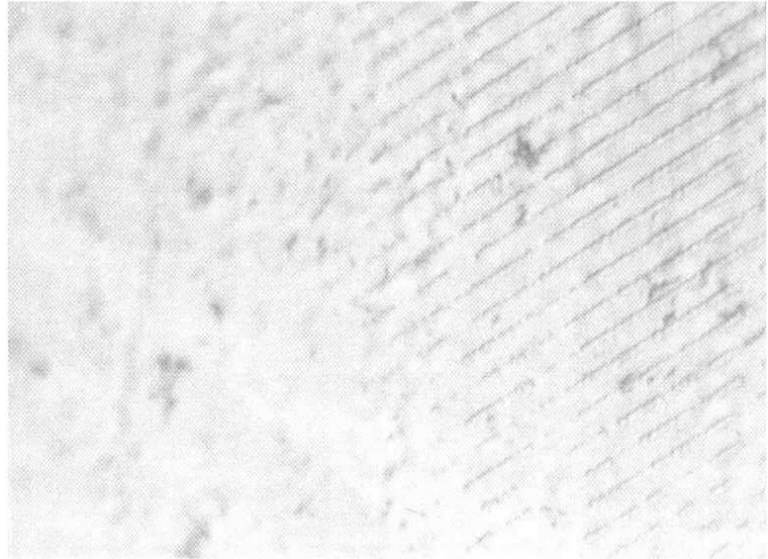}  }
 \caption{\label{FigBarelyPatterned} For a variety of conditions, despite long development times,
 the photoresist AZ5206 would fail to be patterned. }
\end{figure}

To this day the process for 5206 is still not well characterized
by our group.  We began with exposures of 7.2 seconds and ended
with exposures near 4.5 seconds.  In all cases we found similar
problems.  Either the photoresist that patterns the gratings would
loose adhesion leaving a spagetti like structure, or the developer
would leave a layer near the bottom of the grating grooves that
blocked the ICP from etching a grating at all.  In all cases, we
used HMDS.  An example of loss of adhesion is shown in figure
\ref{FigSpagettiPR}.  An example of the photoresist being barely
patterned is shown in figure \ref{FigBarelyPatterned}.


We did find generally better behavior as we lowered the exposure
time.  Both symptoms still occurred (sometimes on the same
exposure), but with lower frequency. We progressed by reworking
the actual sample about 5 times until we had a yield near 70\% of
the pixels.

The pixel quality could be judged quickly by the color at the
lowest microscope magnification. If all pixels were red, then
without fail, the PR would be patterned perfectly upon closer
inspection.  If the pixels were multicolored, we generally found
poor gratings upon inspection with higher magnifications.

One possible suggestion is to try an anti-reflective coating (ARC)
below the photoresist.

The second major issue with the gratings is the etch rate.  Using
two data points, we estimated the etch rate of the grating region
using the CHTM ICP H2Anneal recipe to be about 2400 Angstroms per
minute and the etch rate through the $Si_3N_4$ to be 920 Angstroms
per minute. However, the etch rate for mask GDS1 and GDS2 are
different because they have different groove widths.  The etch
rate of the gratings should be better characterized.

\section{Removing the substrate}

The task of removing the substrate required two practice rounds
before we tried the last set of adjustments on the actual device.
As described in the previous chapter, we used a 4:1 citric acid to
$H_2O_2$ wet etch.

The etching was performed in a 1000 mL beaker with about 500 mL of
etching solution in the beaker.  A magnetic stirring rod was used
to improve the etch rate and etch uniformity.  The etch rate is in
the range of 0.2 microns per min to 0.4 microns per min.  This
gives a total etch time of nearly 36 hours.

During the etching, I would remove the samples from the etching
solution, rinse and dry the samples, and measure their thickness
with a micrometer.

Each time, before placing the samples in the etch solution, one
should dip the samples into a solution of 1:10 $NH_4OH$ to $H_2O$
to remove the surface oxide.  In principle, this should improve
the etch rate uniformity across the sample.

I first practiced the etching on some samples mounted to glass
slides with photoresist.  I found the etch rate slowed with time.
After 4 to 6 hours the etching solution needed to be replaced.
During the last 10 hours of the etch, the photoresist broke free
from the glass slide and the samples were etched away from both
sides.

The next iteration involved mounting the sample to glass slides
with melted wax.  This time I was testing two different samples: a
brag mirror with many 3000 Angstrom layers of AlAs, and a
quantum-well structure with a single 300-Angstrom etch-stop layer
of AlAs. When the sample was less than 10 microns thick, the brag
mirror crumbled.  I suspect the crumbling was due to the stress of
the oxide formed in the reaction with the AlAs
layer\cite{97CarterComan01}.

On the quantum-well sample, I saw the rough surface turn into a
mirror-like surface over the course of four hours.  The
mirror-like surfaces would start as small spots at random
locations on the surface and grow.  When the surface was 90 \%
mirror-like, I removed it from solution.  Visually the surface
looked smooth with small discolorations.  An alpha step scan
revealed that the etch stop layer had broken through and the
etching had again paused on the AlGaAs forming the barriers of the
quantum wells.

We needed to improve the etch rate uniformity or decrease the
amount of wet etching necessary to remove the substrate. Our
solution was to lap 400 microns from the substrate before
performing the wet etch.  With this additional step, we removed
the substrate trivially.  The surface turned from rough to
mirror-like in less than 5 minutes.

The substrate removal revealed a few indications of the
wafer-fusion quality.  The substrate removal revealed that the
windows buried between the two epilayers had warped the epilayer
above it into a 0.3 micron arroyo. Second, during a buffered oxide
etch (BOE) dip to remove the AlAs etch-stop layer, the epilayer
above one pixel popped off.

\section{Etching the $Si_3N_4$}

Three approaches were considered for etching the $Si_3N_4$.  First
was an ICP chlorine-based etch.  This etches GaAs twice as fast as
the $Si_3N_4$.   Although a small etch into the GaAs would not be
disastrous for the process, the combined uncertainties of our etch
rate in the ICP made this less desirable.

A second possibility is to use a BOE etch.  I tried using a 90
second 1:20 BOE etch on three occasions on practice wafers.  On
two occasion, the BOE etched the large contact strips without
difficulty, but did not etch the small 15 micron holes on the
middle mesa that are needed to make contact with this layer. The
third occasion will be discussed in the next paragraph.

\begin{figure}
 \centerline{  \includegraphics[width=4in]{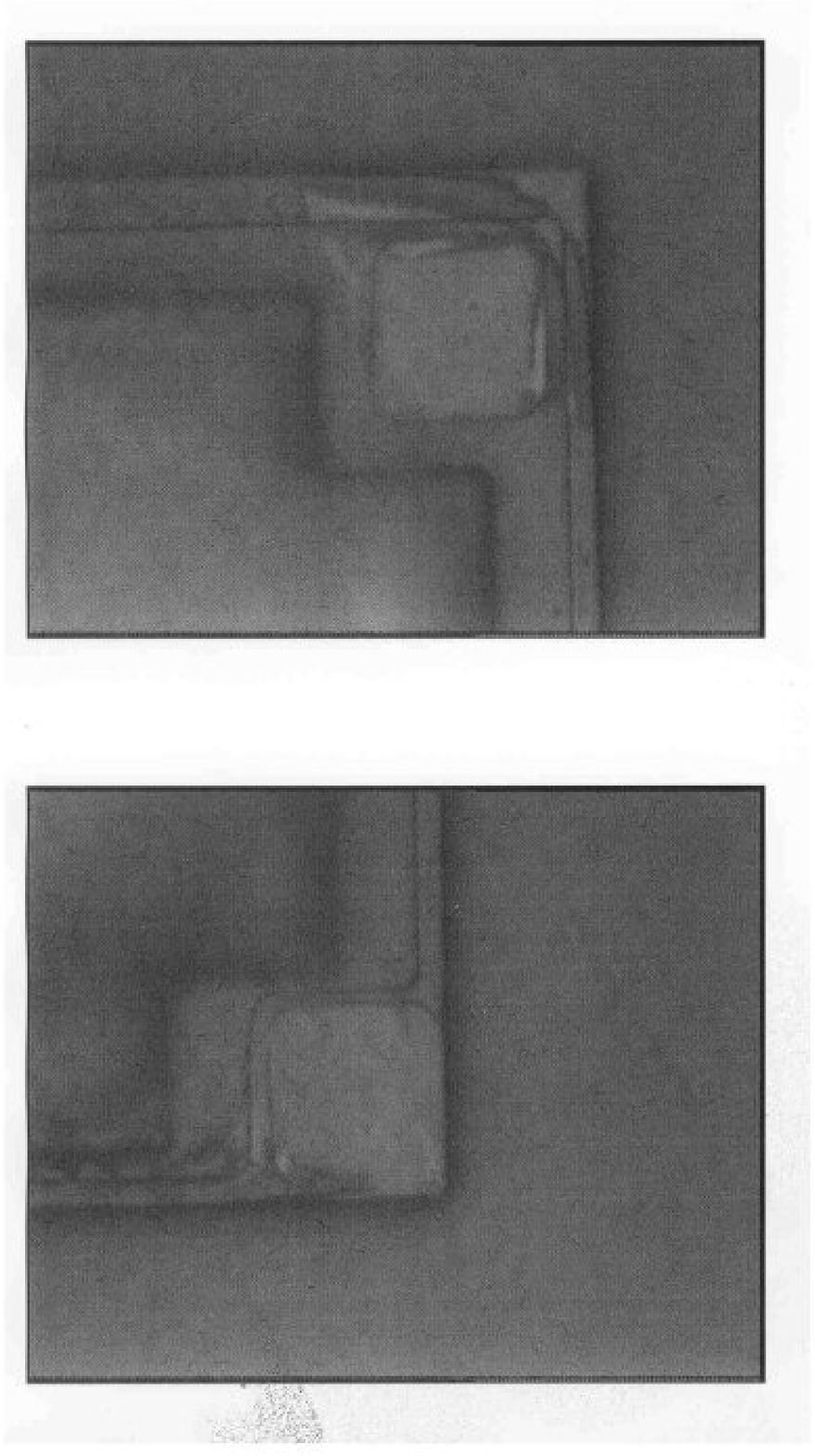}  }
 \caption{\label{FigSiNEtch} An optical microscope image showing the $Si_3N_4$ etched to the mesa wall. }
\end{figure}

The final method considered was a Reactive Ion Etch (RIE) with
$CHF_3$.  This process worked beautifully on three trial samples.
For a 100 watt plasma, and 100 mTorr of $CHF_3$, the etch rate
appeared to be more than 300 Angstroms per min.  When we placed
our actual, offical sample in the RIE, we found the process was
etching more strongly on the edges near the mesa walls.  Figure
\ref{FigSiNEtch} shows the $Si_3N_4$ etched away to the walls. We
believed the etching on these edges was due to residual $O_2$
attacking the number of sharp corners in the photoresist near this
corner.

On the third re-work of this step, we tried purging the $O_2$ from
the RIE exensively.  During the conditioning runs, with the main
$O_2$ valve behind the RIE shut off, we opened the $O_2$ line in
the RIE to evacuate the pipe of all $O_2$. We also did a 30 minute
conditioning run. Before the etch, we performed 8 pump-and-purge
cycles to remove the atmospheric $O_2$.

During the actual etch, we ran out of $CHF_3$.  Under a
microscope, the etch rate was observed to be highly non-uniform,
but the edges near the mesa walls were untouched.  Because we did
not have an option to get more $CHF_3$ for several weeks, we
choose to finish the etch with a one-minute etch in 1:20 BOE.
About 80\% of the $Si_3N_4$ appeared to have cleared, but about
20\% was still a brown or blue color.  After much concern, we
proceeded with the process for depositing the bottom and middle
contacts.

As we discuss later in this chapter, the device exhibited
open-circuit behavior on all devices.  $Si_3N_4$ will form an
insulating barrier for 5 volts with as little as 50 Angstroms
\cite{91Mui01}. For a thickness of 50 Angstroms, the $Si_3N_4$
will appear the same color as the semiconductor.  We could have a
small layer of $Si_3N_4$ insulating our substrate from our metal
contacts.

To use $O_2$ or not to use $O_2$, that is the question. The
Reactive Ion Etching (RIE) of the $Si_3N_4$ was done with a
$CHF_3$ plasma.  Other users use a small amount of $O_2$ in the
RIE plasma. The reason is the $CHF_3$ etching process tends to
leave behind organic compounds \footnote{I would like to thank
Capt Chris Morath at AFRL/SN and Dr. Gabriel Font-Rodriguez at the
US Air Force Academy for bringing this concern to my attention.}.
The small amount of $O_2$ is intended to remove the organic
compounds from the surface.  As shown in figure \ref{FigSiNEtch},
adding $O_2$ to the $CHF_3$ during the RIE etch will not work for
our process. One possible solution is to first perform the $CHF_3$
etch, then after purging the system and removing the photoresist,
to do an $O_2$ plasma clean.

\section{Driving metal up the mesa walls}

The original plan was to use a wet etch for the bottom mesa. The
reason was to create a slopped side wall that would facilitate the
metal climbing up the mesa wall to make contact with the middle
layer. The wet etch was supposed to be nearly isotropic to make
these slopped side walls.

We tried several variations of a wet etch in $H_3PO_4:H_2O_2:H_2O$
of concentration 1:1:45.  We tried with and without a magnetic
stirring rod and we tried a stronger concentration of 1:1:15.  The
etch rate was very non-isotropic.  The solution appeared to be
etching laterally a factor of 40 times faster than the etch rate
downward.

Instead of the wet etch, we decided to do a straight wall ICP etch
and mount the sample in the metal evaporator at a small angle. The
samples were mounted on the rough side of a Si wafer and loaded at
an angle of 5.5 degrees such that the wall that the metal was
going to climb was facing the oncoming particle flux.

\begin{figure}
 \centerline{  \includegraphics[width=4in]{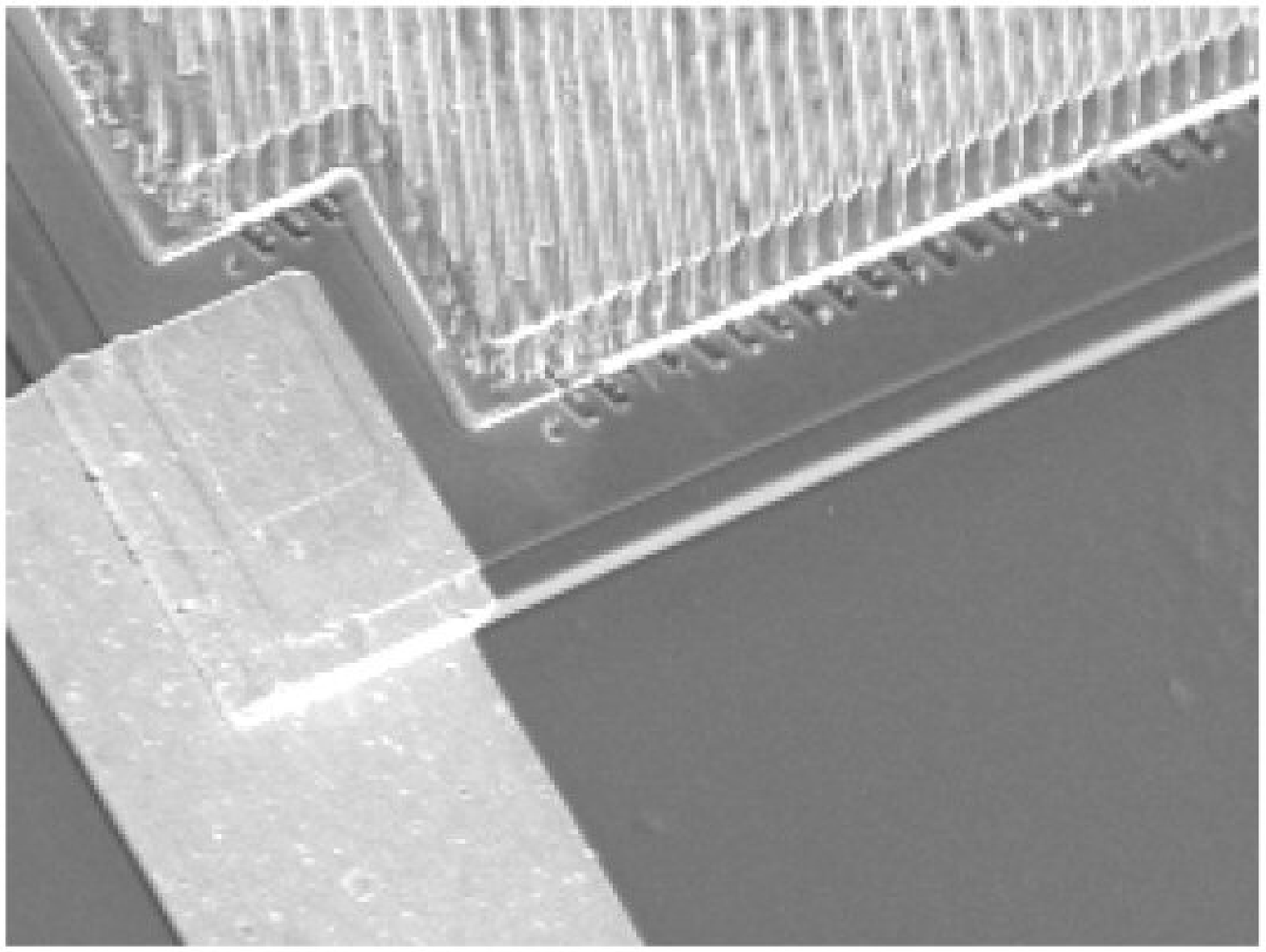}  }
 \caption{\label{FigMetalMesaWall} An SEM showing metal deposited on the mesa walls. }
\end{figure}

This technique has shown mixed success.  Figure
\ref{FigMetalMesaWall} shows our success in getting the metal to
climb the wall.  However, on all three iterations that we tried
this process, the lift-off procedure was difficult.  In two cases,
the lift-off procedure also removed metal that was intended to
stick to the surface, and left metal where it was suppose to be
removed. On one iteration, the lift-off procedure completely
removed all the metal leaving nothing.

I think the cause for these complications is the metal-flux angle
is overshooting the angle of the undercut of the photoresist.
Metal lift-off is best performed with image reversal photoresist.
The image reversal leaves the photoresist with an undercut that
forces clean breaks in the metal deposited on the surface from the
metal deposited on the photoresist.  If during the metal
deposition, the sample is slopped at an angle higher than the
undercut, then the metal will climb up the photoresist walls and
tie the two layers together.

To verify that the angle was responsible for the difficulties, we
processed other samples in parallel with our sample.  For the
other samples that were loaded flat in the metal evaporator, the
lift-off procedure was smooth and uneventful.  The metal was all
lifted-off within about 2 minutes using the acetone spray brush.
For the polarimeter loaded at a 5.5 degree angle during the same
run, lift-off involved more than 30 minutes with the acetone spray
brush.  The resulting metal deposited was partly a success but
also had many regions of metal lifted-off that were suppose to
stay and many regions of the surface with metal that was suppose
to be lifted off.

Adhesion to the $Si_3N_4$ was also considered a potential source
of difficulty. On the final metal evaporator run with the official
sample, we tried using a 50 Angstrom layer of titanium to improve
the metal adhesion to $Si_3N_4$.

\section{Wire bonding}

Once the metal was annealed, we proceeded to wire bond the
structure.  The middle-layer contact pads required care to bond
to.  If the wrong temperature was used, the metal would rip from
the $Si_3N_4$ insulating it from the bottom contact layer.
Patience was the key for bonding these structures.

\begin{figure}
 \centerline{  \includegraphics[width=4.5in]{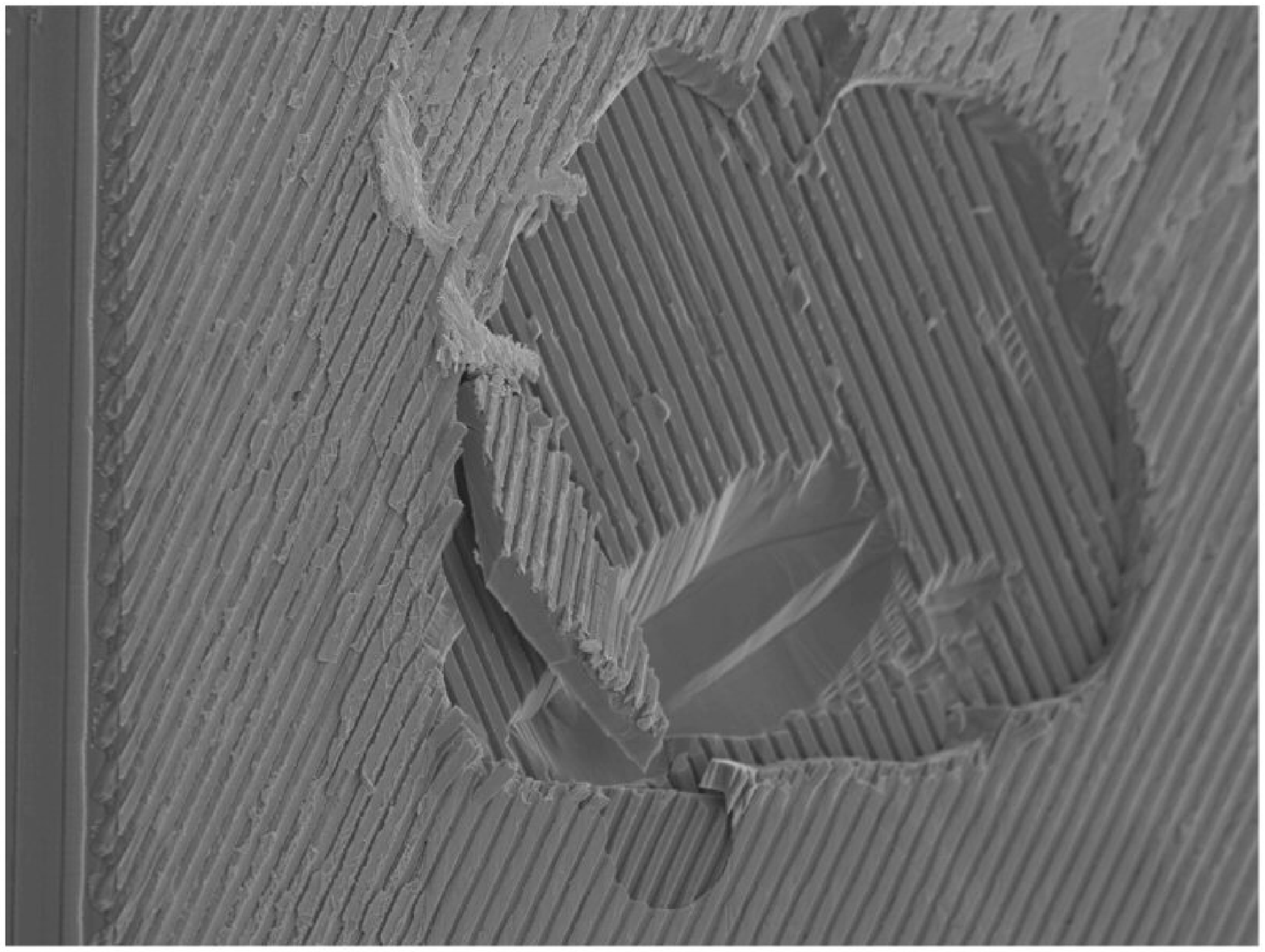}  }
 \caption{\label{FigWireBondingDamage} An SEM showing how the wire bonding pulled the fused epilayer from the top of the pixel. }
\end{figure}

The top contacts were more difficult.  Each time a wire bond was
made to the top surface, the bond would break loose, and the
contacting metal surface would disappear.  Closer inspection with
an SEM revealed the source of the difficulty.  Figure
\ref{FigWireBondingDamage} shows one such example of a pixel where
the attempt to bond to the top contact failed.
Every top-contact to which a traditional wire-bond was attempted
resulted in the failure of the fused interface.  The wire bonds
were adhering to the top metal contact pads without difficulty,
but they were breaking off the fused epilayer.

Sunil Raghavan was able to make contact to the pixels despite this
complication.  The solution involved a conducting epoxy and can be
seen in figure \ref{FigSEMContactPads}.

\section{The failure of ohmic contacts}

After the device was wire bonded, we tested the structure with a
semiconductor parameter analyzer.  Every pair of contacts,
including the ground to ground contacts, showed an open behavior
at room temperature with resistances greater than 1 giga ohm.  A
typical QWIP device at room temperature has a resistance near 10
to 20 ohms.

Where we had multiple contacts on a single ground strip, we would
get the expected short like behavior with resistance less than 1
ohm.  This verified the set-up and wiring diagrams were correct.
The resistance of a quantum-well structure rises as one lowers the
temperature.

To test the quality of the ohmic contacts, we used the ohmic
contact probing mesas.  Probes were placed on the top contact
probing mesa. The current did not need to do anything different
from entering the n-doped layer and leaving the n-doped layer 100
microns away. The resistance was still greater than a giga ohm.
Unfortunately the middle and bottom ohmic contact probing mesas
did not have any metal due to the difficulties during the metal
lift-off procedure.

There are several possible causes for the failure to form ohmic
contacts.

One possible cause is $Si_3N_4$ remained on the surface due to the
difficulties we had removing it. $Si_3N_4$ will form an insulating
barrier for 5 volts with as little as 50 Angstroms \cite{91Mui01}.
For a thickness of 50 Angstroms, the $Si_3N_4$ will appear the
same color as the semiconductor.  We could have a small layer of
$Si_3N_4$ insulating our substrate from our metal contacts.

Another possibility is the RIE etch left a small layer of organics
on the surface that is insulating the metal contacts from the
n-doped surface.  If this is the case, no amount of $CHF_3$
etching will remove this layer without use of some $O_2$. Perhaps
after the $Si_3N_4$ etch with $CHF_3$, a very  pure $O_2$ plasma
will ensure the surface is free of an organic insulator.  Pure
$O_2$ plasma should not attack the $Si_3N_4$ so long as there is
no residual $CHF_3$.

A third possibility is the $1.75 \times 10^{18}\ cm^{-3}$ Si
doping is not enough to form good ohmic contacts.   To evaluate
this, we should process a much more simple structure and evaluate
the quality of the ohmic contacts.


\section{Characterization setup}

Also as a part of the project, at AFRL/VS, we validated an
approach to characterize the polarimeter.  A wire grid polarizer
was inserted in-between a monochometer and the dewar window.  A
known QWIP sample with a linear grating was placed in the dewar.
We measured the photocurrent as a function of polarizer angle.
This experiment showed that the photocurrent was related to the
angle of the polarizer in complete agreement with theory.

\section{Suggestions for future processing rounds}

The good news: at the end of the first round of processing, the
physical structure on the chips is the desired physical structure.
Figures \ref{FigSEMContactPads}, \ref{FigMetalMesaWall} and
\ref{FigWireBondingDamage} show two quantum-well stacks and two
gratings with one grating is buried in between the two
quantum-well stacks.  This is itself a substantial accomplishment.

For the next round of processing, I have the following
suggestions.

 \begin{enumerate}
  \item Test the quantum-wells and the contact layer of the wafer with a simple
  structure.
  \item Characterize the photoresist AZ5206 better or try a
  different photoresist.  Some suggested an anti-reflective
  coating could help.
  \item Better characterize the etch rate of the gratings.
  \item Test for residue left from the RIE etch of the $Si_3N_4$.  Build a test structure with an n-doped GaAs surface.
  Deposit $Si_3N_4$.  Etch the $Si_3N_4$ holes, and deposit metal.  Test for
  conduction through the n-doped GaAs.
  \item Test if the metal climbs the walls without the lift-off
  complications at a smaller angle (3 degrees for example).
\end{enumerate}


\chapter{Conclusions}

This thesis reported on a new optoelectronic device to detect,
simultaneously and instantaneously, the four parameters that fully
describe the polarization state of an ensemble of incident photons
in a narrow wavelength band.

In chapter \ref{ChapPolBackground}, I introduced the Stokes
parameters as a way of fully describing the polarization state of
incident light.  I also introduced polarimetric imaging, and I
reviewed other techniques for capturing polarimetric images. Other
techniques involve large errors at the edges of objects in the
scene or require the designer to sacrifice spectral resolution or
increase the weight by using beam splitters and multiple focal
plane arrays.

Chapter \ref{ChapConceptOfOperation} introduced my solution to
this difficulty.  The new device uses the interference between
multiple diffractions and reflections in the structure to encode
in the multiple photocurrents the polarimetric information.

A computer model, developed in chapter \ref{ChapModeling},
demonstrates the device's preliminary polarimetric capabilities.
The chapter describes the methodology of the model, reviews data
generated by the model, and analyzes the data to gather worst-case
performance estimates.

Finally, in chapters \ref{ChapProcessDesign} and
\ref{ChapLessons}, I explain our approach and progress in building
the device.  These sections reviewed the overall approach, the
wafer-fusion process, the substrate removal process, and recounted
the difficulties we had in building the device:  the difficulty
with etching the gratings, the difficulties with etching the
$Si_3N_4$, and the difficulty with the metal lift-off.  The final
product did not have ohmic contacts into the device.  I also
listed a few possible explanations for this failure: an insulator
was left on the surface or the contact layer's doping is too low.
Appendix \ref{SecProcessingStepsDetailed} will enumerate the
detailed processing recipe that I suggest for future fabrication
attempts.

The polarimeter-in-a-pixel will be a totally new type of
optoelectronic sensor.  This is the first sensor that detects all
the polarization information on a single physical pixel.  The
first round of fabrication brought into focus many hurdles that we
overcame in building the physical structure.  I expect that after
solving the problem with the ohmic contacts, the concept for the
device will be successfully validated.

\chapter*{Appendices}

\appendix

\chapter{Suggested processing recipe}
\label{SecProcessingStepsDetailed}

Procedure for processing the Polarimeter-in-a-pixel:

\begin{figure}
 \centerline{  \includegraphics[width=4.5in]{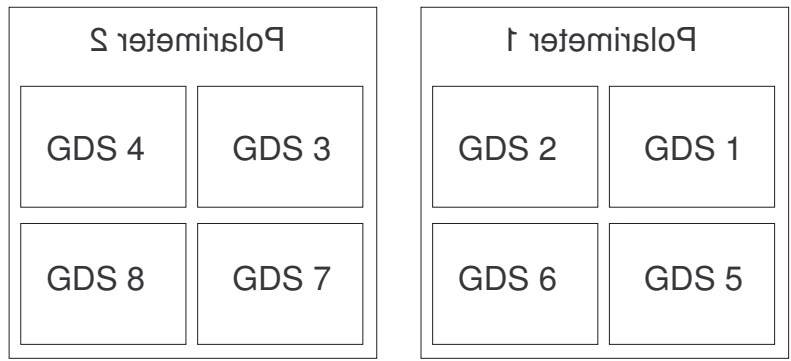}  }
 \caption{\label{FigMaskLayout} The location of the various masks on the mask set. The designated locations
 are placed in perspective by noticing the mirror image of the words `Polarimeter 1' or `Polarimeter 2' on the mask and on this figure. }
\end{figure}

\begin{enumerate}

\item  Prepare the surface for wafer fusion.

Clean surface of tweezers, and both wafers. This is the most
critical step for successful wafer fusion.  A 5-minute acetone
soak (to remove organics), A 5-minute isopropanol soak (to remove
acetone), A 5-minute soak in running DI water (to remove
isopropanol residue). Dip into 1:5 HF for 30 seconds to remove
native oxides formed, then a 5-minute soak in running DI water to
remove the HF and allow safe handling of the wafer.  Now a 20
second dip in 1:30 NH4OH : H2O. Immediately blow dry (removes last
of oxides) and  place in PECVD vacuum in less than 10 minutes to
beat oxide re-growth.

PECVD: Deposit 2000 Angstroms of Si3N4 oxide on both wafers to
protect the clean surface.  Use a surface temperature = 300 C, an
operating pressure of 500 mTorr, use N2 flow = 85 sccm, NH3 flow =
12 sccm, SiH4 flow = 65 sccm, RF Power = 55 watts. It should take
about 17 minutes for 2000 Angstroms to grow.  Let the sample cool
to 150 C before removing from the chamber.

\item   Photolithography (this step is still producing a 70\%
pixel yield, and needs to be further refined). Only one of the two
wafers needs this photolithography.

Clean the $Si_3N_4$ surface. Note humidity and temperature. Again
a 5-minute acetone soak and an acetone spray brush (to remove
organics). Swirl the acetone constantly. Next, a 5-minute
isopropanol soak (to remove acetone), again swirling constantly.
Now a 5-minute soak in running DI water (to remove isopropanol
residue). Blow dry the sample.

Dehydration Bake:  150 C for 7 min. Cover surface with HMDS, spin
at 5k for 30 seconds. Bake on hotplate 150 C for 30 seconds.  Add
positive PR AZ 5206 to the wafer. Beth recommended this PR because
it produces a more thin coat to get sharp grating features. Spin
at 5k, 30 seconds.   Check for photoresist that climbed over edge
onto bottom.  If found, restart step 2.

Softbake 90 C for 90 seconds. Remove edge beads (optional). If you
do remove edge beads, use the edge bead removal mask. Expose
between 3.5 and 5.7 seconds. Develop for ~80 seconds in 1:5 AZ400
developer. Use the Dielectric-Grating mask (a negative mask)
located on GDS-1.  See figure \ref{FigMaskLayout} for more mask
location.

\item   ICP Etch of the dielectric grating through the $Si_3N_4$.

We need a reproducible etch depth known depth to within 0.02
microns. Our two-point calibration showed that using the H2Aneal
recipe the etch rate in $Si_3N_4$ is 926 Angstroms  per min.  For
2000 Angstroms, the etch time is 2.1587 min.  After etching the
$Si_3N_4$ one needs to etch the GaAs.  The etch rate is about 2410
Angstroms per min. With a desired depth of 8800 Angstroms, the
etch time is an additional 3.65 min.

Total etch time is 5.81 min = 5 min 49 seconds.

Next we remove the photoresist with an acetone spray brush,
followed by standard cleaning.

Standard cleaning is a 5-minute acetone soak and spray (to remove
organics). Next a 5-minute isopropanol soak (to remove acetone).
Now a 5-minute soak in running DI water (to remove isopropanol
residue). Blow dry.

You will probably need to use an $O_2$ plasma to remove the
remaining photoresist from surface.  After the $O_2$ plasma,
repeat the above cleaning procedure with soaks in acetone, IPA, DI
water, and blow dry.

\item   Wafer fuse the samples.

In the case of this thesis, we shipped the samples to Lincoln Lab
Attention Dr. Z. Liau MIT Lincoln Laboratory 244 Wood Street,
Lexington, MA 02420-9108.  He is doing this as a first-time small
effort favor, and will probably not be able to perform wafer
fusion on a regular basis without some funding support.

Wafer Fusion begins with removing Si3N4 protective layer with BOE
wet etch.  Use about 1 minute and 30 seconds of BOE 1:20 to remove
the 2000 Angstroms of $Si_3N_4$.

In this run, the wafer fusion was performed for 16 hours under 550
C and pressure.

\item   Lapping substrate of wafer down to about 75 to 100
microns.

\item  Remove substrate from wafer 2.

The substrate to be removed is from the wafer which has never been
etched. Measure sample thickness with an electronic micrometer.

Perform a standard clean. Next dip in NH4OH : H2O 1:10 for 45
seconds to remove surface oxides, rinse in H2O and Blow dry. Mount
the sample on glass slide with wax.  Place side of wafer not to be
etched closest to the glass.  Mounting on the slide is performed
by heating a slide up to 112 C; melt wax on slide; place wafer
with protected substrate face down onto the wax; slide around
until one can tell no bubbles remaining (look through bottom).
After wax cools, measure thickness of sample on the slide - Mix by
weight 1:1 dry Citric Acid Monohydrate and deionized water (~ 250
grams, 250 grams) - Mix by volume 4:1 Citric Acid : $H_2O_2$.
Place a magnetic stirring stick under a small platform in beaker.
Pour 4:1 solution into the beaker. Etch rate is some where 0.25 to
0.35 microns per min, monitor thickness throughout etch and adjust
endpoint accordingly.  Replace the solution every 2 to 4 hours
because the H2$O_2$ will lose its potency. Last, remove the
300-Angstrom AlAs etch stop layer with a 60 second etch in 30:1
BOE solution.

\item  Photolithography

Use standard recipe for CHTM cleanroom to add 5214-IR Photoresist
to newly exposed episurface. Expose using the Reflective-Grating
(negative mask)-GDS-2 mask pattern. See figure \ref{FigMaskLayout}
for mask location.  Alignment windows on the buried structure
should be visible in the microscope.  Develop with standard time.

\item  ICP etch (etch rate needs to be recalibrated)

Make grooves 0.9 microns deep.  Using the previously found grating
etch rate for the H2Anneal recipie, a 9000 Angstroms etch at 2400
Angstroms per min =  3 min 45  seconds.

Remove the photoresist by acetone spray brush, followed by a
standard clean.  An $O_2$ plasma clean will probably be needed.

\item  Photolithography.

 Add 5214-IR photoresist. Use the imager-reversal to optimize
lift-off. Becomes a negative photoresist.  Expose TopMetal
(positive mask)-GDS-3 mask. See figure \ref{FigMaskLayout} for
mask location. Develop.

\item Deposit n-Metal.  To aid in making contacts, use a very thick deposition.
at least 4000 Angstroms if not more.  Lift-off in with acetone
spray brush.

\item  Photolithography

Add 5214 photoresist.  Use the standard CHTM recipe for positive
tone behavior. Expose TopPixelMesa (positive mask)-GDS-4 mask. See
figure \ref{FigMaskLayout} for mask location. Develop.

\item ICP etch of pixel mesa.

PR covering Au with little extra serves as mask (4.565 microns =
(4.865-0.3) microns, etch slowly to get right stopping point) Etch
rate 2700 Angstroms per min  = 16 min 54 seconds.

Remove the photoresist with aceton spray brush.  $O_2$ plasma will
probably be needed.

\item Photolithography

Add 5214-IR photoresist. Use standard CHTM image reversal recipe
to make it a negative photoresist.  Expose UpperMiddlePixelMesa
(negative mask) GDS-5 mask.  See figure \ref{FigMaskLayout} for
mask location. Develop.

\item ICP

Etch 1.0 micron - this will bring us 0.7 microns deep into the 0.9
micron gratings. @ 2700 Angstroms per min = 3 min 42 sec

Remove the photoresist with acetone spray brush.  Standard clean.
You will probably need to use an $O_2$ plasma to clean the
remaining photoresist.

\item Photolithography

Add 5214-IR photoresist using standard CHTM image reversal recipe.
Expose LowerMiddlePixelMessa (negative mask) GDS-6 mask. See
figure \ref{FigMaskLayout} for mask location.  Develop.

\item. ICP etch GaAs.

Etch 4.065 microns  = 4.865-0.3 - 0.5. The first -0.7 is for the
0.7 microns of the bottom wafer etched by the 2nd ICP etch(this
etch took away 1 micron, 0.3 from top wafer, 0.7 from bottom
wafer).  The next -0.5 microns is so that we stop short of the
substrate and allow the bottom contact a conductive channel to the
pixel.

At 2700 Angstroms per min this etch will be 15 minutes and 3
seconds.

Remove the photoresist, standard clean followed by an $O_2$ plasma
to remove remaining photoresist.

\item PECVD.

Deposit 2000 Angstroms of Si3N4 over whole surface. Use same
recipie as in step 1.  The $Si_3N_4$ dielectric breakdown field is
at 8 MV/cm \cite{91Mui01}. To allow bias of up to 10 V on 2nd QW,
we need thickness greater than 125 Angstroms.  The remainder is a
factor of saftey and to ensure sidewall coverage.

\item Photolithography.

Apply 5214-IR photoresist. Use image reveral to make it a negative
Photoresist. Again use standard CHTM recipe. This type of
photoresist allowed me to make the mask design with clear view of
the pixels with which to align the mask.  Expose with
$Si_3N_4$-Removal-mask (positive mask)-GDS-7 mask.  See figure
\ref{FigMaskLayout} for mask location. Develop  This will exposes
the patern that will open holes in the $Si_3N_4$ on the second
layer, holes for the bottom contact, and holes in pixel top over
the metal.

\item RIE Selective Etch of  Si3N4.

First clear out the $O_2$ in the pipe. Bring the RIE to about 10
mTorr.   Now make sure the $O_2$ valve behind the RIE is closed,
and turn on the $O_2$ switch on the RIE.  Wait till the pressure
returns to about 10 mTorr.  Now close the $O_2$ swith on the RIE
and perform a long conditioning run with CHF3.  About 100 watts
with about 100 mTorr pressure.    30 minutes should be enough. Now
load the sample.  Do eight pump-and-purge cycles to remove any
$O_2$ from the atmosphere in the chamber.  Now do the CHF3 etch
for 9 minutes with the same settings as the conditioning run.

Now remove the photoresist with acetone spray brush followed by a
standard clean.

Next do an $O_2$ clean. Again do an extensive purge of the system
to remove any CHF3 in the chamber.  Clear out the CHF3 pipe just
as we cleared out the $O_2$ pipe in the previous paragraph. Do a
30 minute conditioning run.  Now load the sample, do 4
pump-and-purge cycles, and do a 5 minute $O_2$ plasma clean.  The
$O_2$ plasma (if free of any CHF3) should not attack the $Si_3N_4$
and will remove only the organics left on the surface.

\item Photolithography.

Use 5214-IR photoresist.  Use the image reveral recipe so it
becomes becomes a negative photoresist. Expose to LastMetalization
(postive)-GDS-8 mask. See figure \ref{FigMaskLayout} for mask
location.  Develop

\item Modified n-Metal deposition

First deposit use 50 Angstroms of titatium, then continue with the
regular n-metal deposition.   Lift-off using acetone spray brush,
then perform standard cleaning.

\item Put sample in rapid thermal annealer (RTA) to form Ohmic contacts

Treat in a rapid-thermal annealing system at 430   C for 1 min in
the presence of N2/H2 forming gas, for the formation of good
quality top, middle, bottom ohmic contacts.

\item Mount the sample onto a back-side illuminated chip-carrier
and wire bond.

\end{enumerate}

\bibliographystyle{plain}


\end{document}